\documentclass[useAMS,usenatbib]{mnras}

\usepackage{hyperref}

\usepackage{epsfig}
\usepackage{myaasmacros}
\usepackage{amsmath}
\usepackage{amsfonts}
\usepackage{amssymb}
\usepackage{subfig}
\usepackage{comment}
\usepackage{color}

\newcommand{\Msun}{\hbox{M$_\sun$}}
\newcommand{\hii}{\hbox{H{\sc ii}}}
\newcommand{\ha}{\hbox{H$\alpha$}}
\newcommand{\hb}{\hbox{H$\beta$}}
\newcommand{\oiii}{\hbox{[O\,{\sc iii}]$\lambda5007$}}
\newcommand{\oi}{\hbox{[O\,{\sc i}]$\lambda6300$}} 
\newcommand{\nii}{\hbox{[N\,{\sc ii}]$\lambda6584$}}
\newcommand{\siid}{\hbox{[S\,{\sc ii}]$\lambda\lambda6717,6731$}}
\newcommand{\sii}{\hbox{[S\,{\sc ii}]$\lambda6724$}}

\newcommand{\oiiihb}{\hbox{[O\,{\sc iii}]/H$\beta$}}
\newcommand{\niiha}{\hbox{[N\,{\sc ii}]/H$\alpha$}}
\newcommand{\siiha}{\hbox{[S\,{\sc ii}]/H$\alpha$}}
\newcommand{\oiha}{\hbox{[O\,{\sc i}]/H$\alpha$}}

\newcommand{\nv}{\hbox{N\,\textsc{v}\,$\lambda1240$}}

\newcommand{\siliv}{\hbox{Si\,\textsc{iv}\,$\lambda1400$}}
\newcommand{\nivd}{\hbox{[N\,\textsc{iv}]$\lambda1483$+N\,\textsc{iv}]$\lambda1487$}}
\newcommand{\niv}{\hbox{N\,\textsc{iv}]\,$\lambda1485$}}
\newcommand{\civd}{\hbox{C\,\textsc{iv}\,$\lambda\lambda1548,1551$}}
\newcommand{\civ}{\hbox{C\,\textsc{iv}\,$\lambda1550$}}
\newcommand{\heii}{\hbox{He\,\textsc{ii}\,$\lambda1640$}}
\newcommand{\oiiiuvd}{\hbox{O\,\textsc{iii}]$\lambda\lambda1661,1666$}}
\newcommand{\oiiiuv}{\hbox{O\,\textsc{iii}]\,$\lambda1663$}}
\newcommand{\niii}{\hbox{N\,\textsc{iii}]\,$\lambda1750$}}
\newcommand{\silii}{\hbox{Si\,\textsc{ii}]\,$\lambda1814$}}
\newcommand{\siliiid}{\hbox{[Si\,\textsc{iii}]$\lambda1883$+Si\,\textsc{iii}]$\lambda1892$}}
\newcommand{\siliii}{\hbox{Si\,\textsc{iii}]\,$\lambda1888$}}
\newcommand{\ciiid}{\hbox{[C\,\textsc{iii}]$\lambda1907$+C\,\textsc{iii}]$\lambda1909$}}
\newcommand{\ciii}{\hbox{C\,\textsc{iii}]\,$\lambda1908$}}
\newcommand{\cii}{\hbox{C\,\textsc{ii}]\,$\lambda2326$}}
\newcommand{\oiid}{\hbox{[O\,{\sc ii}]$\lambda\lambda3726,3729$}}
\newcommand{\oii}{\hbox{[O\,\textsc{ii}]\,$\lambda3727$}}

\title[Synthetic nebular emission] 
{Synthetic nebular emission from massive galaxies II: ultraviolet-line diagnostics of 
dominant ionizing sources}  
\author[Hirschmann et al.]{Michaela Hirschmann$^{1,2}$\thanks{E-mail:
hirschma@iap.fr}, St\'ephane Charlot$^{1}$, Anna Feltre$^{1,3}$, Thorsten Naab$^{4}$,  \newauthor Rachel
S. Somerville$^{5,6}$,  Ena Choi$^{7}$, \\
$^{1}$Sorbonne Universit\'es, UPMC-CNRS, UMR7095, Institut d'~Astrophysique de Paris, F-75014
Paris, France\\
$^{2}$University of Vienna, Institute for Astronomy, T\"urkenschanzstrasse 17, 1180 Vienna, Austria\\ 
$^{3}$Univ.\,Lyon, Univ.\,Lyon1, ENS de Lyon, CNRS, Centre de Recherche Astrophysique de Lyon, UMR5574, 69230 Saint-Genis-Laval, France\\
$^{4}$Max-Planck-Institute for Astrophysics, Karl-Schwarzschild-Strasse 1, 85741 Garching, Germany \\
$^{5}$Department of Physics and Astronomy, Rutgers, The State
University of New Jersey, NJ 08854, USA \\
$^{6}$Center for Computational Astrophysics, Flatiron Institute, 162 5th Ave, New York, NY 10010, USA \\
$^{7}$Department of Astronomy, Columbia University, New York, NY 10027, USA \\
}

\begin{document}

\date{Accepted ???. Received ??? in original form ???}

\pagerange{\pageref{firstpage}--\pageref{lastpage}} \pubyear{2002}

\maketitle

\label{firstpage}

\begin{abstract}

We compute synthetic optical and ultraviolet (UV) emission-line properties of galaxies in a full cosmological framework by coupling, in post-processing, new-generation nebular-emission models with high-resolution, cosmological zoom-in simulations of massive galaxies. Our self-consistent modelling accounts for nebular emission from young stars and accreting black holes (BHs). We investigate which optical- and UV-line diagnostic diagrams can best help to discern between the main ionizing sources, as traced by the ratio of BH accretion to star formation rates in model galaxies, over a wide range of redshifts. At low redshift, simulated star-forming galaxies, galaxies dominated by active galactic nuclei and composite galaxies are appropriately differentiated by standard selection criteria in the classical \oiii/\hb\ versus \nii/\ha\ diagram. At redshifts $z\ga1$, however, this optical diagram fails to discriminate between active and inactive galaxies at metallicities below $0.5 Z_\odot$.
To robustly classify the ionizing radiation of such metal-poor galaxies, which dominate in the early Universe, we confirm 3 previous, and propose 11 novel diagnostic diagrams based on equivalent widths and luminosity ratios of UV emission lines, such as EW(\oiiiuv) versus \oiiiuv/\heii, \ciii/\heii\ versus \oiiiuv/\heii, and \civ/\ciii\ versus \ciii/\cii. We formulate associated UV selection criteria and discuss some caveats of our results (e.g., uncertainties in the modelling of the \heii\ line). These UV diagnostic diagrams are potentially important for the interpretation of high-quality spectra of very distant galaxies to be gathered by next-generation telescopes, such as the {\it James Webb Space Telescope}. 

\end{abstract}


\section{Introduction}\label{intro}

The emission from interstellar gas contains important information about the nature of ionizing sources in a galaxy. In particular, optical nebular emission lines are traditionally used to estimate whether ionization is dominated by young massive stars (tracing the star formation rate, hereafter SFR), an active galactic nucleus (hereafter AGN) or evolved, post-asymptotic giant branch (hereafter post-AGB) stars \citep[e.g.,][]{Izotov99, Kobulnicky99, Kauffmann03, Nagao06, Kewley08, Morisset16}. In fact, the intensity ratios of strong emission lines, such as \hb, \oiii, \ha\ and \nii, exhibit well-defined correlations, characteristic of different ionizing sources. One of the most widely used line-ratio diagnostic diagrams, originally defined by \citet[][hereafter BPT]{Baldwin81} and \citet{Veilleux87} in that defined by the \oiii/\hb\ and \nii/\ha\ (hereafter simply \oiiihb\ and \niiha) ratios. This diagram has proven useful to identify the nature of the ionizing radiation in large samples of galaxies in the local Universe \citep[e.g.][]{Kewley01, Kauffmann03}, but its applicability at high redshift is still unclear. Based on a sample of about 50 star-forming galaxies and 10 confirmed AGN at $z\sim2.3$, \citet{Coil15} find that the local AGN/star-forming (SF) galaxy classification in the \oiiihb--\niiha\ diagram robustly separates these populations in the distant Universe. However, the apparent evolution of the \oiiihb\ ratio over cosmic time identified in several observational studies challenges this conclusion \citep[e.g.,][]{Shapley05, Lehnert09, Yabe12, Steidel14, Shapley15, Strom17}. In addition, at the very low metallicities expected in the youngest galaxies at high redshifts \citep[e.g.,][]{Maiolino08}, emission-line ratios for SF- and AGN-dominated models tend to occupy similar regions of the \oiiihb--\niiha\ diagram \citep{Feltre16}. Thus, even if optical emission lines will be measurable out to redshifts of several using the future {\it James Webb Space Telescope} ({\it JWST}), their usefulness to constrain the nature of ionizing sources in the early universe is uncertain. 

In recent years, interest has grown in ultraviolet (UV) nebular emission lines, such as \ciii, \civ\ and \heii\ lines, which can be routinely detected in star-forming galaxies at low redshift with the {\it Hubble Space Telescope} and at higher redshift with near-infrared spectrographs \citep[e.g., ][]{Pettini04, Hainline09, Steidel14, Shapley15, Erb10, Hainline11, Stark14, Stark15, Berg16,Vanzella16, Senchyna17,Talia17, Laporte17}. UV emission lines tend to be particularly prominent in metal-poor, actively star-forming dwarf galaxies at all redshifts \citep[e.g.,][]{Stark14,Berg16,Senchyna17}. This is because metal-poor gas cools less efficiently than metal-rich gas, leading to higher electron temperatures and stronger collisionally excited emission lines, while metal-poor stars are also hotter and have harder ionizing spectra than metal-rich ones \citep[e.g.,][]{Schaller92, Schaerer03}. This makes UV emission lines attractive tracers of the ionizing radiation in metal-poor galaxies, which will be observable with {\it JWST} at redshifts way into the reionization epoch, between $z\sim 15$ \citep{Planck2016} and $z\sim6$ \cite[e.g.][]{Fan2006}. Yet, to date, the dependence of these lines on the nature of the ionizing radiation in galaxies has been far less studied than that of optical lines, and never in a cosmological context.

In a pioneering study, \citet{Feltre16} explored the UV emission-line properties of photoionization models of SF- and AGN-dominated galaxies and proposed new line-ratio diagnostic diagrams to discriminate between the two populations, such as \ciii/\heii\ versus \civ/\heii\ diagram. This approach \citep[see also][]{Nakajima18} relies on a blind exploration of the wide parameter space of photoionization models (ionization parameter, element abundances, depletion of metals on to dust grains, stellar and AGN properties, etc.). While highly instructive, it also presents some drawbacks: the results may be affected by parameter combinations not found in nature; the combined effects of star formation and nuclear activity in a `composite' galaxy were not explored;\footnote{Note that \citet{Nakajima18} consider composite-like galaxies, by simply adopting different fractional AGN contributions, but they do not derive any selection criteria for composites.} and the conclusions drawn about line-ratio diagnostic diagrams do not incorporate the potential evolution of galaxy properties with cosmic time. 

In this context, modelling nebular emission from galaxies {\em in a full cosmological framework} could provide valuable insight into the connection between observed emission lines and the underlying ionizing-source properties as a function of cosmic time. Yet, fully self-consistent models of this kind are currently limited by the performance of cosmological radiation-hydrodynamic simulations and insufficient spatial resolution on scales of individual ionized regions around stars and AGN. As an alternative, some studies proposed the post-processing of cosmological hydrodynamic simulations and semi-analytic models with photoionization models to compute  nebular emission of galaxies in a cosmological context \citep{Kewley13, Orsi14, Shimizu16, Hirschmann17}. For example, by combining chemical enrichment histories from cosmological simulations with photoionization models of SF galaxies, \citet{Kewley13} proposed modified, {\em redshift-evolving} criteria in the \oiiihb--\niiha\ diagram to select SF galaxies while accounting for the predicted cosmic evolution of \oiii/\hb. Despite this progress, no study so far has either considered a full differentiation between SF, composite and AGN-dominated galaxies, or investigated the potential usefulness of UV emission lines for the classification of ionizing sources over cosmic time.

In this paper, we close this gap by appealing to the methodology introduced by \citet{Hirschmann17} to model in a self-consistent way the emission from different gas components ionized by different sources in simulated galaxies. This is achieved by coupling photoionization models for AGN \citep{Feltre16}, young stars \citep{Gutkin16} and post-AGB stars \citep{Hirschmann17} with cosmological hydrodynamic simulations. The resulting models reproduce observations of galaxies in various optical BPT diagrams involving the \hb, \oiii, \oi, \ha, \nii\ and \siid\ (hereafter simply \sii) lines, at both low and high redshifts, and accounting for the observed evolutionary trend in \oiiihb\ \citep{Hirschmann17}. Our methodology, which by design captures SF, composite and AGN-dominated galaxies in a full cosmological framework, provides a unique means of answering key questions we wish to address in the present study:
\begin{itemize}
\item To what extent are optical selection criteria, traditionally used to identify the nature of ionizing sources in low-redshift galaxies, still useful for classifications at high redshift?
\item Can we identify novel diagnostic diagrams in the rest-frame UV and derive corresponding selection criteria to identify the main ionizing sources of galaxies, in particular in the distant Universe?
\end{itemize}
Answers to these questions will allow proper interpretation of the high-quality spectra of very distant galaxies to be gathered by next-generation telescopes, such as {\it JWST}, providing valuable insight into, for example, the relative contributions by star formation and nuclear activity to reionization of the Universe.

The paper is structured as follows. In Section \ref{theory}, we present the general theoretical framework of our study, including the zoom-in simulations of massive galaxies, the nebular-emission models and the way in which we combine the former with the latter. Sections \ref{opticaldiagrams} and \ref{UVdiagrams} describe our main results about tracing the nature of ionizing sources in distant galaxies via standard optical and novel UV line-ratio diagnostics. We address possible caveats of our approach and discuss our findings in the context of previous theoretical studies in Section \ref{discussion}. Finally, Section \ref{summary} summarizes our results.

\section{Theoretical framework}\label{theory} 

\subsection{Cosmological zoom-in simulations of massive haloes}\label{simulations} 

To achieve the analysis presented in this paper, we appeal to a set of 20 high-resolution, cosmological zoom-in simulations of massive haloes  described in \citet{Choi16} and \citet{Hirschmann17}. We briefly summarise these models below and refer the reader to the original studies for more details. 

\subsubsection{The simulation code `SPHGal'}\label{setup}

The simulations used in this study were produced with a modified version of the highly parallel, smoothed particle hydrodynamics (SPH) code \textsc{Gadget3} \citep{Springel05a},  SPHGal. As described in \citet{Hu14}, this code includes `modern' numerical SPH schemes, which pass  all standard tests previously reported to be problematic, such as fluid-mixing problems \citep[see also,][]{Choi16, Nunez17}.

SPHGal follows baryonic processes, such as star formation, chemical enrichment, metal-line cooling, stellar and AGN feedback and ultraviolet photo-ionization background \citep{Haardt01}. Specifically, star formation and chemical evolution are modelled as described in \citet{Aumer13} and \citet{Nunez17}.  To model star formation, we assume a temperature-dependent density threshold, above which gas particles get Jeans unstable and are stochastically converted into star particles. Chemical enrichment is accounted for via type-Ia and type-II supernovae (SNe) and AGB stars, tracing 11 elements (H, He, C, N, O, Ne, Mg, Si, S, Ca and Fe) in both gas and star particles. We also account for metal diffusion in the ISM as in \citet{Aumer13}. 

Star formation is regulated by both stellar and AGN feedback. We adopt the approach outlined in \citet{Nunez17} to model mass, energy and momentum injection due to early stellar and SN feedback from different evolutionary stages of massive stars. AGN feedback is tied to the prescription for BH growth described in \citet{Choi16}, where BHs accrete gas following a statistical Bondi-Hoyle approach \citep{Bondi52, Choi12}. To compute AGN feedback from BH accretion, we do not make the widely used assumption of considering only thermal energy release into the ambient medium \citep[e.g.,][]{Hirschmann14, Naab16, Somerville15}. Instead, we rely on a more physically motivated approach including both mechanical feedback, motivated by broad-absorption-line winds from quasars, and radiative X-ray feedback, due to Compton and photoionization heating and radiation pressure \citep{Ostriker10, Choi16}. We note that our 20 zoom-in simulations do not include any metallicity-dependent heating prescription. This is justified by the fact that, as shown by \citet{Choi16}, such refinements are not found to have any significant impact on basic properties of massive galaxies. Adopting these sub-grid models, in particular the improved prescription for AGN feedback, we obtain fairly realistic massive galaxies, e.g. in terms of star formation histories, baryon conversion efficiencies, sizes, gas fractions, gas and stellar metallicities, hot-gas X-ray luminosities and optical emission line ratios \citep{Choi16, Hirschmann17, Brennan18}.

\subsubsection{The simulation set-up}\label{setup}

Our set of 20 cosmological zoom-in simulations is based on a sub-set of initial conditions from \citet{Oser10, Oser12} and \citet{Hirschmann12, Hirschmann13}, adopting a WMAP3 cosmology ($\sigma_8 = 0.77$, $\Omega_{\mathrm{m}}=0.26$, $\Omega_{\Lambda}=0.74$ and $h= 0.72$; see, e.g., \citealp{Spergel03}). The dark-matter (DM) haloes chosen for zoom-in re-simulations, with $z=0$ virial  masses between  $2.2 \times 10^{12} \Msun\ h^{-1}$ and $2.2 \times 10^{13} \Msun \ h^{-1}$, were selected from  a DM-only N-body simulation with a co-moving periodic box length $L=72\ \mathrm{Mpc}\ h^{-1}$ and $512^3$ particles \citep{Moster10}.  To construct the initial conditions for the high-resolution re-simulations, individual haloes are traced back in time, and all particles closer to the halo centre than twice  the radius where the mean density drops below 200 times the critical
density of the universe at any given snapshot are identified. These DM particles  are replaced with particles at higher resolution with masses of $m_{\mathrm{dm}} = 2.5\times 10^7 \Msun\,h^{-1}$ for DM, and $m_{\mathrm{gas}} = 4.2\times 10^6\Msun\,h^{-1}$ for gas, equal to that of star particles. The co-moving gravitational softening length of the DM particles is $890\ h^{-1}\mathrm{pc}$, and that of the gas and star  particles $400\ h^{-1} \mathrm{pc}$.  

To investigate the redshift evolution of different galaxy properties (including emission-line ratios), we construct stellar merger trees for  the sample of 20 model galaxies described in the previous subsection. As in \citet{Oser12}, we start by using a friends-of-friends algorithm to identify, at any simulation snapshot, a central galaxy -- the host (i.e., the most massive  galaxy sitting at the minimum of the halo potential well) -- and its surrounding  satellite (less massive) galaxies. We require a minimum of 20 stellar particles  (i.e., a minimum mass of about $1.2 \times 10^8 \Msun$) to identify a galaxy.  At $z = 2$, all galaxies in our sample are more massive than about $10^{10}\Msun$, implying that, at $z<2$, we resolve mergers down to a mass ratio of at least $1:100$. In the analysis presented in the remainder of this paper, we trace back at every time step only the most massive progenitor of a present-day galaxy, i.e., we focus on central galaxies. 

Note that our limited sample of zoom-in simulations does not allow for a statistical representation of galaxies (e.g., probability distribution), and moreover, our simulation suite probes only a subset of the parameter space (e.g., both lower-mass and higher-mass halos are missing). Nevertheless, we emphasise that  despite using a non-cosmologically representative sample for this study, our simulations {\it do} follow the scaling relations between various physical parameters (e.g. mass-metallicity relation).

\subsection{Modeling of nebular emission}\label{nelms} 

As in \citet{Hirschmann17}, we post-process the re-simulations of 20 galaxies presented in Section~\ref{simulations} to include nebular emission. To achieve this, we adopt the recent prescriptions of \citet{Gutkin16},  \citet{Feltre16} and \citet{Hirschmann17}, with some minor modifications, to compute the nebular emission arising from young massive stars, narrow-line regions of AGN and post-AGB stars, as described in Section~\ref{cloudymodels} below. All emission-line models  presented in this paper were computed using version c13.03 of the photoionization code \textsc{Cloudy} (\citealp{Ferland13}). Then, we couple these extensive calculations of nebular-emission models  with the simulations of massive galaxies, as described in Section~\ref{coupling}. For further details about the nebular-emission models and coupling methodology, we refer the reader to \citet{Gutkin16}, \citet{Feltre16} and \citet{Hirschmann17}.

\subsubsection{Nebular-emission models}\label{cloudymodels}

To model \hii\ regions around young stars, we adopt the updated grid of nebular-emission models of star-forming galaxies computed by J.~Gutkin (private communication). As in \citet{Gutkin16}, these calculations combine the latest version of the \citet{Bruzual03} stellar population synthesis model (Charlot \& Bruzual, in preparation) with \textsc{Cloudy}, following the method outlined by \citet{Charlot01}. The models used here include newly updated spectra of Wolf-Rayet stars from the Potsdam Wolf-Rayet Models (PoWR; private communication from H. Todt). The resulting grid encompasses models in wide ranges of interstellar (i.e. gas+dust-phase) metallicity, $Z_{\star}$, ionization parameter, $U_{\star}$, dust-to-metal mass ratio, $\xi_d$, \hii-region density,  $n_{\mathrm{H}, \star}$ and carbon-to-oxygen abundance ratio, (C/O)$_{\star}$ (see table~1 of \citealp{Hirschmann17} for a summary of all parameters). We adopt here the default emission-line predictions of \citet{Gutkin16} for 10\,Myr-old stellar populations with constant SFR and a standard \citet{Chabrier03} initial mass function (IMF; consistent with the IMF adopted in the simulations), truncated at 0.1 and 300 $\Msun$.\footnote{We adopt here an IMF truncated at $m_{\rm up}=300\Msun$, rather than the standard 100 $\Msun$, to increase the strength of the \heii\ line and bring it in better agreement with observations \citep{Senchyna17}. See also section \ref{HeII} for further discussion.} 

For narrow-line regions of AGN, we adopt an updated grid of nebular-emission models of \citet{Feltre16}. The new grid includes microturbulent clouds (with a microturbulence velocity of 100~km/s) and adopts a smaller inner radius (90~pc compared to the old radius of 300~pc for an AGN luminosity of $10^{45}$~erg/s) for the gas in the NLR.\footnote{Note that these modifications have been found to result in a better agreement with the observations of NV emission lines of AGN  (Mignolie, Feltre et al. in prep.).} In this prescription, the spectrum of  an AGN is approximated by a broken power law of adjustable index $\alpha$ in the frequency range of ionizing photons \citep[equation 5 of][]{Feltre16}.  The grid of AGN nebular-emission models is parametrized in terms of the interstellar metallicity in the narrow-line region, $Z_{\bullet}$, the ionization parameter of this gas, $U_{\bullet}$, the dust-to-metal  mass ratio, $\xi_d$, the density of gas clouds, $n_{\mathrm{H}, \bullet}$, and the carbon-to-oxygen  abundance ratio, (C/O)$_\bullet$ (see table~1 in \citealp{Hirschmann17}). 

To describe nebular emission from quiescent, passively evolving galaxies, we adopt the  grid of  `PAGB' models introduced in \citet{Hirschmann17}, constructed by inserting spectra of single-age, evolved stellar populations into the photoionization code \textsc{Cloudy}. The grid includes models in wide ranges of stellar population age and metallicity, $Z_{\diamond, {\rm stars}}$, gas ionization parameter, $U_{\diamond}$, dust-to-metal mass ratio, $\xi_{\rm d}$, hydrogen density, $n_{\mathrm{H}, \diamond}$ and interstellar metallicity, Z$_{\diamond}$ (see  table~1 of \citet{Hirschmann17}). 

\subsubsection{Coupling nebular-emission models with zoom-in simulations}\label{coupling} 

We couple the extensive grid of nebular-emission models described in Section~\ref{cloudymodels} with the simulations of massive galaxies described in Section~\ref{simulations} by selecting a SF, AGN and PAGB emission-line model for each simulated galaxy at each redshift step. The sum of these three components makes up the integrated nebular emission of a model galaxy.  In practice, we select the SF/AGN/PAGB models appropriate for each galaxy by self-consistently matching all model parameters possibly available from the simulations (e.g., metallicity of the star-forming gas). 

A few model parameters cannot be retrieved from the simulation, such as the slope of the ionizing AGN spectrum $\alpha$, the dust-to-metal mass ratio $\xi_{\rm d}$, and the hydrogen gas density in individual ionized regions $n_{\mathrm{H}}$. By considering different values for these parameters, their potential impact on the nebular emission of galaxies is {\em automatically} accounted for in our analysis. Specifically, we sample two values of the dust-to-metal mass ratio, $\xi_{\rm d}=0.3$ and $\xi_{\rm d}=0.5$,  in the SF, AGN and PAGB models, as these values are closest to that of $\xi_{\rm d,\odot}=0.36$ in the Solar neighbourhood \citep{Gutkin16}. We further consider two values of the hydrogen gas density in HII-regions, $n_{\mathrm{H}, \star} =10^2\,$cm$^{-3}$ and $n_{\mathrm{H}, \star} =10^3\,$cm$^{-3}$,  in the SF models. In the AGN models, we allow for two values of the gas density in narrow-line regions, $n_{\mathrm{H},\bullet} = 10^3\,$cm$^{-3}$ and $n_{\mathrm{H},\bullet} = 10^4\,$cm$^{-3}$, and two values for the UV-slope of the AGN ionizing spectrum, $\alpha = -1.4$ and $\alpha = -2.0$.\footnote{In other words, we consider all permutations of the non-constrained parameters to have the maximum possible area spanned by line ratios of a galaxy in the BPT/UV diagnostic diagrams.} For the PAGB models, $n_{\mathrm{H}, \diamond} = 10\,$cm$^{-3}$ is adopted. In the next paragraphs, we briefly describe the methodology to couple the SF, AGN and PAGB nebular models with our zoom-in galaxy simulations.   

To match the SF models with simulated galaxies,  we associate the SF emission-line model from the \citet{Gutkin16} grid with the  gas and star parameters closest to that of each galaxy at each simulation time step. Specifically, we select the grid metallicity $Z_{\star}$, carbon-to-oxygen ratio (C/O)$_{\star}$ and  ionization parameter $\log U_{\star}$  closest to the simulated global (i.e. galaxy-wide) metallicity $Z_{\mathrm{gas, glob}}$, abundance ratio (C/O)$_{\mathrm{gas, glob}}$  of the warm-gas phase and the simulated ionization parameter $\log U_{\mathrm{sim}, \star}$. This uniquely defines the \citet{Gutkin16} model associated with each simulated galaxy at each time step. Note that we compute  the ionization parameter of a simulated galaxy using equation~1 of \citet{Hirschmann17}. As explained in \citet{Hirschmann17}, in our approach,   $U_{\mathrm{sim},   \star}$ depends on the simulated SFR (via the rate of ionizing photons) and global average gas density, $\rho_{\mathrm{gas, glob}}$ (via the filling factor).

To associate AGN models from the \citet{Feltre16} grid with the nuclear activity of simulated galaxies  at any simulation time step, we adopt a procedure similar to that for SF models. We take the ISM conditions for the AGN model to be the {\it central} (rather than global) ones  of the simulated galaxy, i.e, in a co-moving sphere of 1-kpc radius around the black hole. This size should be roughly appropriate to probe the narrow-line regions around AGN with luminosities in the range found in our simulations (see fig. 3 of \citealp{Hainline14} and the model AGN luminosities in Fig. 6 of \citealp{Hirschmann17}).  We compute the central warm-gas metallicity, $Z_{\mathrm{gas, 1kpc}}$, central carbon-to-oxygen ratio, (C/O)$_{\mathrm{gas, 1kpc}}$, and central ionization parameter, $U_{\mathrm{sim},\bullet}$, derived from the central volume-averaged gas density, $\rho_{\mathrm{gas, 1kpc}}$, and the bolometric AGN luminosity, $L_{\rm AGN}$ (see \citealp{Hirschmann17} for details).  Then, we select the \citet{Feltre16} model with closest $Z_{\bullet}$, $\log U_{\bullet}$ and (C/O)$_\bullet$.  


\begin{figure*}
\centering
\textbf{All galaxies}\par\medskip\vspace{-0.2cm}
\epsfig{file=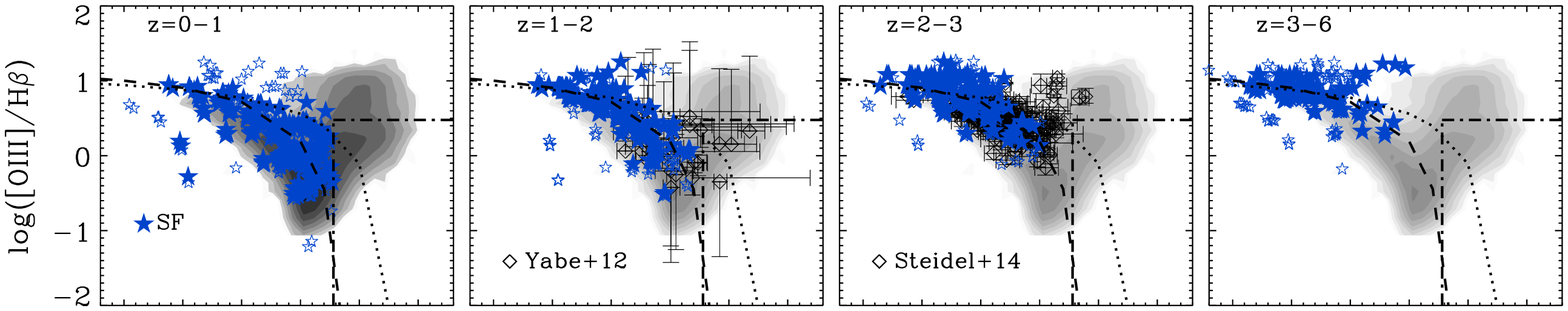,
  width=0.95\textwidth}\\\vspace{-1.2cm}
\epsfig{file=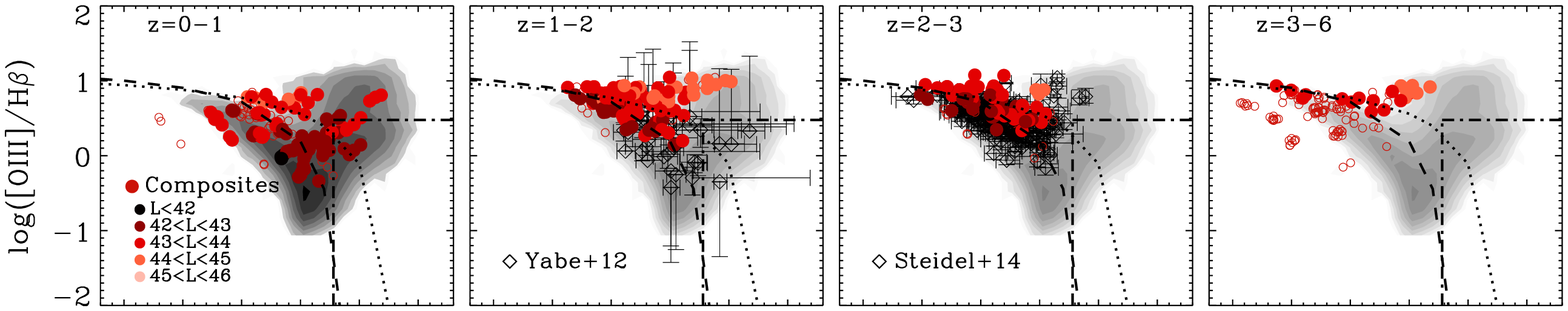,
  width=0.95\textwidth}\\\vspace{-1.2cm}
\epsfig{file=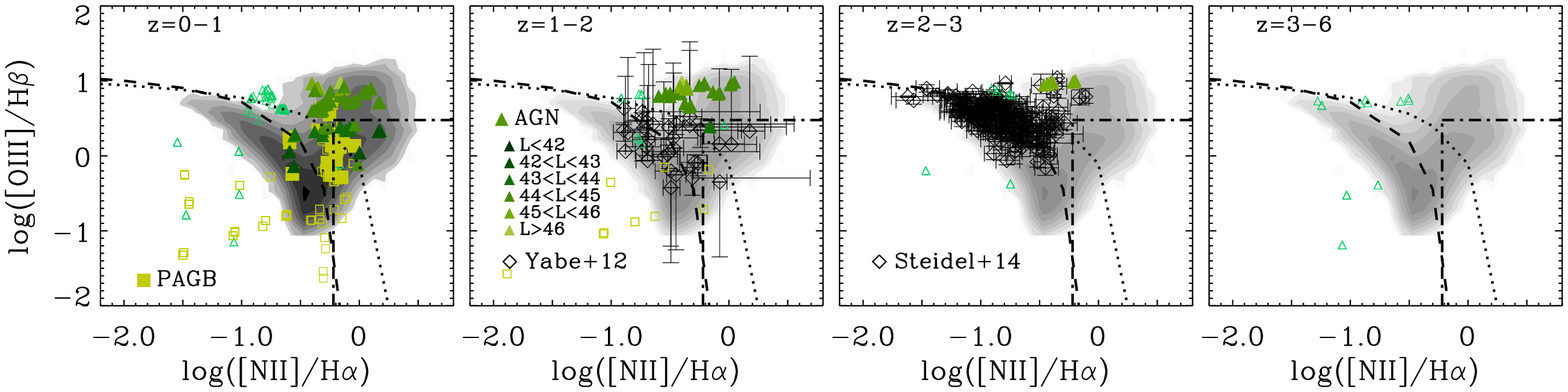,
  width=0.95\textwidth}\\
   \caption{\oiiihb\ versus \niiha\ diagrams for all simulated galaxies and their main high-redshift progenitors in different redshift intervals (different columns). For completeness, the line ratios of simulated galaxies are shown for two different values of four undetermined parameters ($\xi_d=0.3,\,0.5$; $m_{up}=100,\,300 M_\odot$; $n_{H,\star}=10^2,\,10^3 \mathrm{cm}^{-1}$; $\alpha = -1.4,\,-2$). Different coloured symbols in different rows refer to different galaxy types (blue stars in top row: SF; red circles in middle row: composite; green triangles in bottom row: AGN-dominated; and yellow squares in bottom row: post-AGB-dominated galaxies, see text for theoretical distinction criteria). Different AGN luminosities in composite and AGN-dominated galaxies are colour-coded, as indicated. Also shown for reference are observations of local SDSS galaxies (grey shaded areas and contours),  as well as of distant galaxies (black diamonds) by \citep[][$z\sim$1.4]{Yabe12} and \citep[][$z\sim$2.3]{Steidel14}, together with standard observational criteria to distinguish SF galaxies (below the dashed line) from composites (between the dashed and dotted lines), AGN (above the dotted line) and LI(N)ER (in the bottom-right quadrant defined by dotted-dashed lines), according to \citet[][dotted line]{Kewley01}  and \citet[][dashed and dotted-dashed lines]{Kauffmann03}. The small open symbols show the synthetic line ratios of all galaxies, regardless of luminosity, while the large filled symbols show galaxies above a flux limit of $5 \times 10^{-17}$ erg s$^{-1}$ cm$^{-2}$ in all lines.} \label{BPTevol}   
\end{figure*}

To pick a PAGB emission-line model from the \citet{Hirschmann17} grid for each galaxy at each simulation time step, we compute the average age and metallicity of all star particles older than 3\,Gyr, and we adopt the same global interstellar metallicity $Z_{\mathrm{gas,glob}}$, abundance ratio (C/O)$_{\mathrm{gas, glob}}$ and  volume-averaged gas density $\rho_{\mathrm{gas, glob}}$ (used to compute the ionization parameter of the gas ionized by post-AGB stars $U_{\mathrm{sim},\diamond}$) as for \hii\ regions. We then select the PAGB model with closest grid values of $Z_{\diamond, \mathrm{stars}}$, $Age_{\diamond, \mathrm{stars}}$,  $Z_{\diamond}$, (C/O)$_\diamond$, and $\log U_{\diamond}$.

\subsubsection{Total emission-line luminosities and line ratios of simulated galaxies}

The procedure described in the previous paragraphs allows us to compute the contributions of young stars, AGN and post-AGB stars to the luminosities of various emission lines (such as $L_{\mathrm{H}\alpha}$, $L_{\mathrm{H}\beta}$,  $L_{\mathrm{OIII}}$, etc.) in a simulated galaxy. The {\it total} emission-line luminosities of the galaxy can then be calculated by summing over these three contributions. For line luminosity ratios, we adopt for simplicity the notation $L_{\mathrm{OIII}}/L_{\mathrm{H}\beta}=\oiiihb$. In this study, we focus on exploring  line ratios built from six optical lines, 
\hb, \oiii, \oi, \ha, \nii\ and \sii\ (as defined in Section~\ref{intro}),
and 11 UV lines, \nv\ (multiplet), \nivd\ (hereafter simply \niv), \civd\ (hereafter simply \civ), \heii, \oiiiuvd\ (hereafter simply \oiiiuv), \niii\ (multiplet), \silii\ (multiplet), \siliiid\ (hereafter simply \siliii), \ciiid\ (hereafter simply \ciii), \cii\ and \oiid\ (hereafter simply \oii).


\subsubsection{Total equivalent widths of nebular emission lines of simulated galaxies}\label{eqwidth}

In addition to line luminosities and line ratios, we also compute the total equivalent widths (EW) of some nebular emission lines. We obtain the EW of an emission line by dividing the total line luminosity by the total continuum flux C at the line wavelength, e.g. EW(\ciii) $= L_{\rm CIII}/C_{\rm CIII}$ (expressed in \AA). The total continuum C is the sum of the contributions by the stellar (SF and PAGB) and AGN components. For the stellar components, we account for both attenuated stellar radiation and nebular recombination continuum. For the AGN component, we consider only the nebular recombination continuum, and do not account for any attenuated radiation from the accreting BH. This assumption should be reasonable for type-2 AGN, where direct AGN radiation is obscured by the surrounding torus, and only emission from the narrow-line region is observed. Instead, for type-1 AGN, the EW computed here should be interpreted as upper limits. In the remainder of this study, we will focus on exploring the EW of five UV lines: EW(\ciii), EW(\civ), EW(\oiii), EW(\siliii) and EW(\niii). 

\section{Differentiating ionizing sources of galaxies in optical-line diagnostic diagrams}\label{opticaldiagrams} 

\begin{figure}
\centering
\epsfig{file=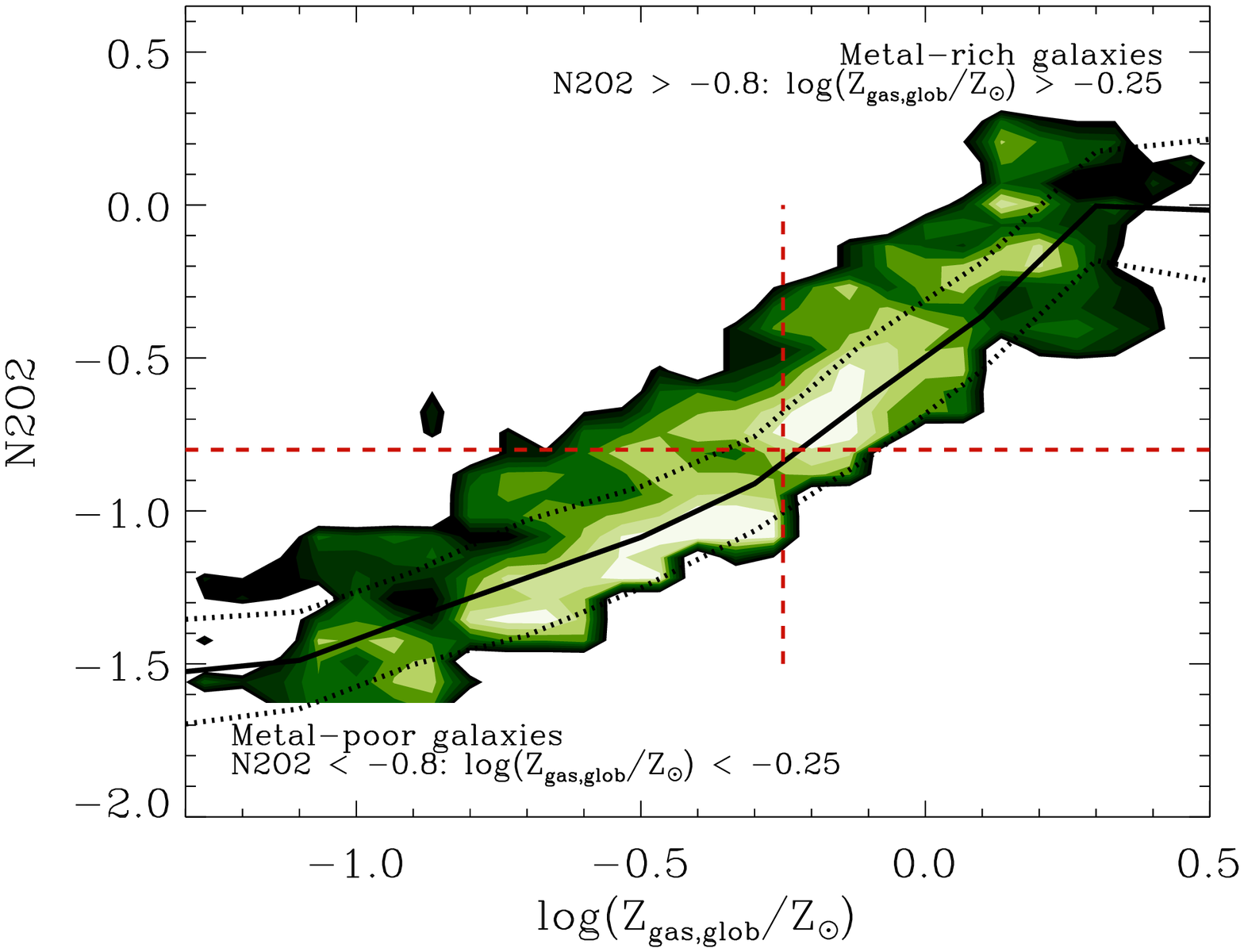, width=0.4\textwidth}
\epsfig{file=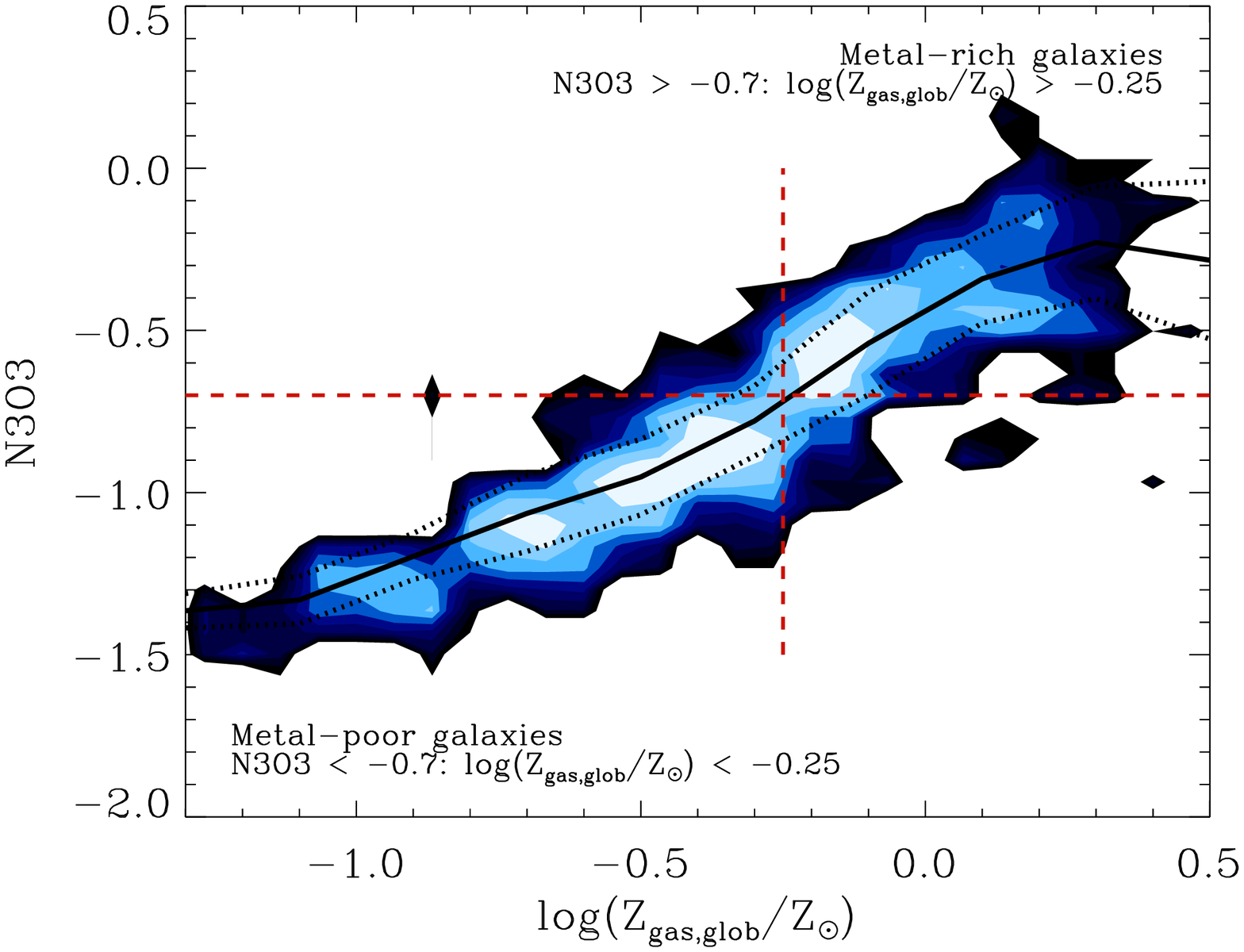, width=0.4\textwidth}
\caption{2D-Histograms of the optical line ratios N2O2 [$\equiv\log(\nii/\oii)$, top panel] and N3O3 [$\equiv\log(\niii/\oiiiuv)$, bottom panel] versus interstellar metallicity, for all simulated galaxies and their main progenitors at $z=0$--6. In each panel, black solid and dotted lines show the mean line ratios and their $1\sigma$ dispersion. The horizontal red dashed line in the top panel indicates the line-ratio cut (N2O2 $=-0.8$) to distinguish between metal-poor and metal-rich galaxies adopted in this work, roughly corresponding to a cut at half-solar interstellar metallicity of $\log(Z_{\mathrm{gas,glob}}/Z_\odot)=-0.25$ (shown by vertical red dashed line). A similar distinction can be achieved with a cut at N3O3 $=-0.7$.}\label{N2O2}     
\end{figure}

\begin{figure*}
\centering
  \textbf{Metal-rich galaxies}\par\medskip\vspace{-0.2cm}
  \epsfig{file=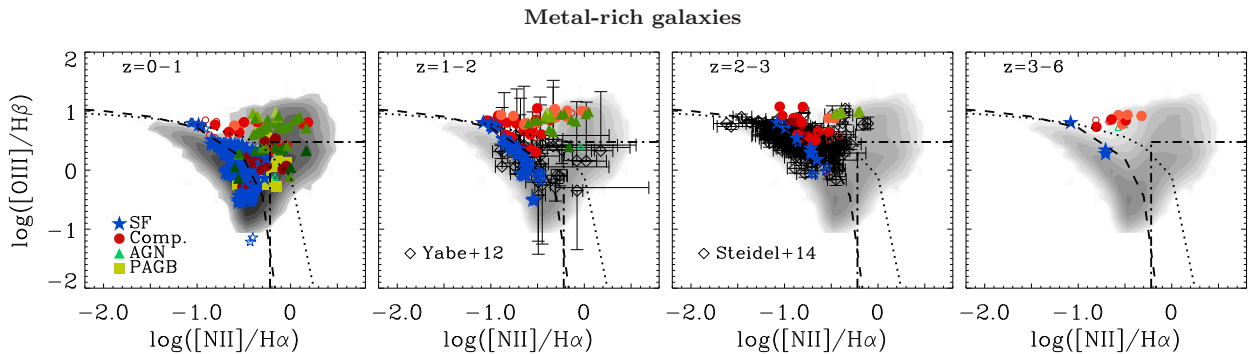,
  width=0.95\textwidth}
   \caption{Same as Fig. \ref{BPTevol}, but for metal-rich galaxies (${\rm N2O2}>-0.8$), and including all galaxy types in a single diagnostic diagram in each redshift interval. } \label{BPTevol_metalrich}   
\end{figure*}

\begin{figure}
\centering
\vspace{-0.5cm}
\epsfig{file=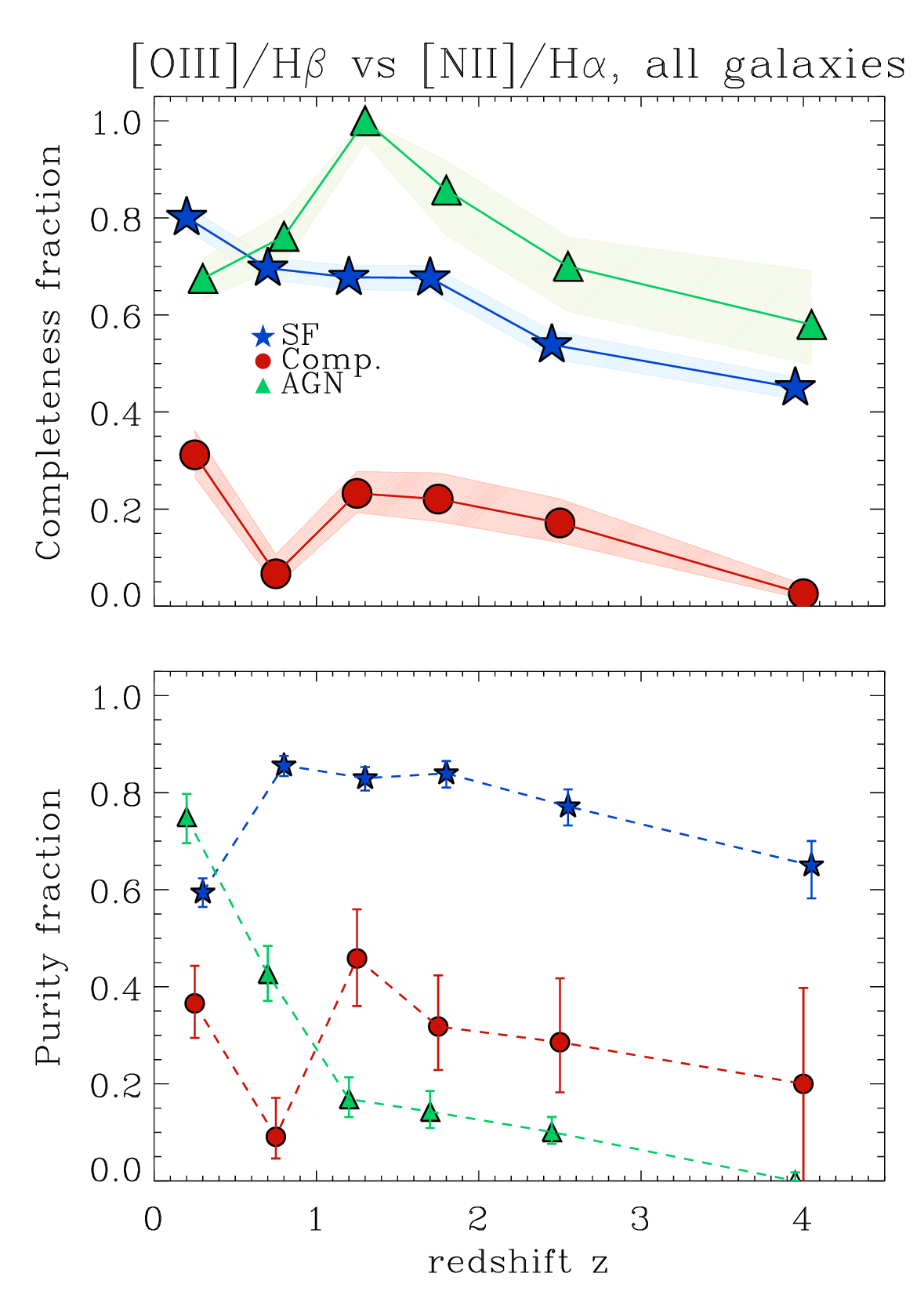, width=0.4\textwidth}\vspace{-0.3cm}
\epsfig{file=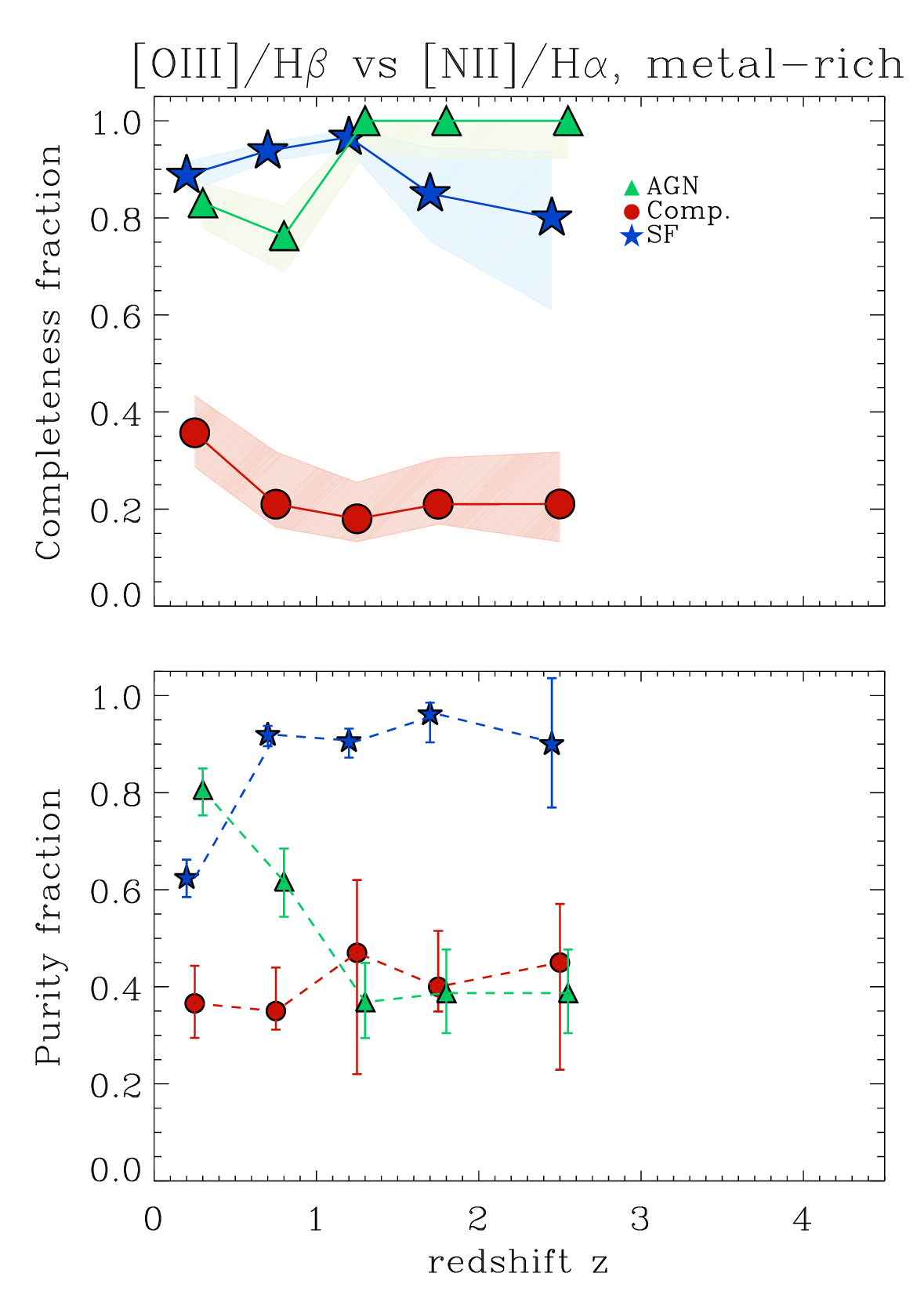, width=0.4\textwidth}
\caption{ Completeness (large symbols, solid lines in the first and third panels) and purity fractions (small symbols, dashed lines in the second and fourth panels) as a function of redshift, for simulated SF-dominated (blue stars and lines), composite (red circles and lines) and AGN-dominated (green triangles and lines) galaxies, as classified observationally using the standard criteria of \citet{Kewley01} and \citet{Kauffmann03} in the optical \oiiihb--\niiha\ line-ratio diagram. The top two panels show the results when including all simulated galaxies, and the bottom two panels when including only galaxies preselected to be metal-rich, with interstellar metallicities $\log(Z_{\mathrm{gas,glob}})\ga-0.25$, using the criterion ${\rm N2O2}>-0.8$ (bottom panel). Error bars and shaded areas illustrate binomial errors. See Section~\ref{opticalfractions} for more details.}\label{OpticalFrac}     
\end{figure}

In this section, we aim to understand to what extent traditionally used selection criteria in the \oiiihb--\niiha\ line-ratio diagram \citep{Kewley01, Kauffmann03} still allow for a differentiation of the main ionizing sources in  {\em distant} galaxies. To robustly assess this, we investigate the full set of 20 zoom-in simulations of massive galaxies and their main progenitors in the optical \oiiihb-\niiha\ diagram in different redshift intervals.\footnote{We do not discuss here the \oiiihb--\siiha\ and \oiiihb-\oiha\ diagnostic diagrams also proposed by BPT, as these are less useful to separate AGN from composite galaxies.}  We consider redshift bins including several simulation snapshots, and hence, potentially several tens of emission-line galaxies. To account for uncertainties arising from the few parameters undetermined by simulations ($n_{\rm H}$, $\xi_d$ and $\alpha$), we allow these parameters to vary as described in Section~\ref{coupling}. Thus, a galaxy at a given time step can appear several times in the line-ratio diagram. 

In the simulations, we can theoretically distinguish between different galaxy types on the basis of the predicted ratio of BH accretion rate (BHAR) to star formation rate (SFR) and the H$\beta$-line luminosity. Specifically, SF-dominated, composite, AGN-dominated and PAGB-dominated galaxies are defined as follows \citep{Hirschmann17}:
\begin{itemize}
 \item SF-dominated galaxies: BHAR/SFR $< 10^{-4}$ and H$\beta_{\rm SF+AGN} > $ H$\beta_{\rm PAGB}$ (blue stars in all figures)
\item Composite galaxies: $10^{-4}<$ BHAR/SFR  $< 10^{-2}$ and H$\beta_{\rm SF+AGN} > $ H$\beta_{\rm PAGB}$ (red circles in all figures)
\item AGN-dominated galaxies: BHAR/SFR $> 10^{-2}$ and H$\beta_{\rm SF+AGN} > $ H$\beta_{\rm PAGB}$ (green triangles in all figures)
\item PAGB-dominated galaxies: H$\beta_{\rm SF+AGN} < $ H$\beta_{\rm PAGB}$ (yellow squares in all figures)
\end{itemize}
In the following subsections, we adopt these theoretical definitions to investigate the locations of different galaxy types in the \oiiihb--\niiha\ diagram from $z=0$ to $z=6$ (Section~\ref{allgals}), discuss possible alternatives when the main ionizing sources in distant galaxies are misclassified (Section~\ref{metalrich}), and quantify the accuracy of traditionally used, optical selection criteria over cosmic time (Section~\ref{opticalfractions}). 

\subsection{Optical line ratios of \textit{all} galaxies}\label{allgals} 

In the left column of Fig.~\ref{BPTevol}, we show the locations of SF-dominated (blue stars, top panel), composite (red circles, middle panel), AGN-dominated (green triangles, bottom panel) and PAGB-dominated (yellow squares, bottom panel) galaxies and their main progenitors extracted from all simulation snapshots at redshifts $z<1$ in the \oiiihb--\niiha\ diagnostic diagram. For composite and AGN-dominated galaxies, the symbols are further colour-coded according to AGN luminosity (as indicated). The grey shaded areas and contours indicate the location of SDSS galaxies in each diagram. 

To perform  a meaningful comparison between models and observations, we show the effect of requiring a typical flux detection limit of $5 \times 10^{17}$ erg s$^{-1}$ cm$^{-2}$  \citep[motivated by table 1 of][]{Juneau14} for all simulated galaxies (as in figure~1 of \citealp{Hirschmann17}). Filled symbols correspond to galaxies satisfying this criterion, and open symbols to those too faint to be detected. Also shown as dashed, dotted and dot-dashed lines in these diagrams are standard observational criteria from \citet{Kewley01} and \citet{Kauffmann03} to distinguish SF galaxies from composites, AGN and LI(N)ER [low-ionization (nuclear) emission] galaxies. Observationally, galaxies with line ratios below the dashed line are classified as SF-dominated,  those with line ratios between the dashed and dotted lines as composite, and  those with line ratios above the dotted line as AGN-dominated. In addition, galaxies with line ratios in the bottom-right quadrant defined by dot-dashed lines are classified as LI(N)ER, whose main ionizing sources are still debated \citep[e.g., faint AGN, post-AGB stellar populations, shocks or a mix of these sources; e.g.,][]{Belfiore16}.

As already discussed in \citet{Hirschmann17}, Fig.~\ref{BPTevol} shows that, at $z<1$, simulated galaxies satisfying our conservative flux limit occupy the same areas as SDSS galaxies in the \oiiihb--\niiha\ plane. Moreover, in general, simulated galaxies of SF, AGN and PAGB types appear to fall in regions of the diagram corresponding to the observationally defined SF, AGN and LI(N)ER categories. Only composite galaxies appear to be distributed more widely than the observations, extending to higher-than-observed \oiiihb\ ratios at the highest AGN luminosities. Nonetheless, the overall agreement between models and observations in Fig.~\ref{BPTevol} is remarkable given that, in our approach, different galaxy types are connected to physical parameters, such as the BHAR/SFR ratio and the fraction of total H$\beta$ luminosity.

Turning towards higher redshifts, the second, third and fourth columns of Fig.~\ref{BPTevol} show the analogue of the first column for the redshift bins $z=1$--2, 2--3, and 3--5, respectively. Similarly to the situation at low redshift, the emission-line properties of simulated galaxies brighter than the flux-detection limit (filled symbols) are consistent with the observed properties of SF galaxies from the samples of \citealp{Yabe12} at $z \sim 1.4$  (black diamonds with error bars in the second column) and Steidel et al. (2014) at $z \sim 2.3$ (black diamonds with error bars in the third column). Interestingly, this agreement arises from the fact that for the whole sample of simulated galaxies, \oiiihb\ globally increases and \niiha\ decreases from low to high redshift. This can be traced back to a drop in interstellar metallicity (the models including secondary N production) and rise in SFR (controlling the ionization parameter) in the models, as discussed in detail in \citet[][]{Hirschmann17}. This makes the different galaxy types less distinguishable toward high redshift in Fig.~\ref{BPTevol}, implying that the traditional optical selection criteria break down. We note that, since the main reason for composite/AGN galaxies to move toward the SF region in the \oiiihb--\niiha\ plane is the lower typical metallicity of high-redshift compared to low-redshift galaxies, the validity of standard optical selection criteria might be preserved by focusing on preselected subsamples of {\em metal-rich} galaxies at any redshift. We will test this hypothesis in the next subsection.

\subsection{Optical line ratios of only \textit{metal-rich} galaxies}\label{metalrich} 

To assess whether traditional, optical selection criteria are useful when applied to metal-rich galaxies at any redshift, we first need to understand how to best distinguish between metal-rich and metal-poor galaxies using observed emission lines. To this goal, we have explored how reasonably different emission-line ratios often used in literature \citep[see e.g.,][and references therein]{Belfiore17, Wuyts16} can trace interstellar metallicity in our simulated galaxies. Here, we report our results for the two line ratios we find to best scale with interstellar metallicity. Specifically, Fig.~\ref{N2O2} shows the fairly tight positive relations between both N2O2 [$\equiv\log(\nii/\oii)$, top panel] and N3O3 [$\equiv\log(\niii/\oiiiuv)$, bottom panel] with interstellar metallicity, for all simulated galaxies and their progenitors at $z=0$--6 (colour-coded by number density; average line ratios with their $1\sigma$ dispersion are shown by solid and dotted lines, respectively).  Note that the primary physical origin of the scatter at fixed metallicity is the dispersion in C/O abundance ratio and dust-to-metal mass ratio $\xi_d$. These tight relations are, to some extent, a consequence of the observationally found relation between N/O and O/H, which is {\em a-priori adopted} in our nebular emission line models \citep[see][for details]{Gutkin16}. Fig.~\ref{N2O2} shows that galaxies with ${\rm N2O2} > -0.8$ or ${\rm N3O3}> -0.7$ (horizontal red dashed lines) have on average interstellar metallicities above approximately half solar [$\log(Z_{\rm gas, glob}/Z_\odot) \ga -0.25$]. 

Fig. \ref{BPTevol_metalrich} shows the analogue of Fig. \ref{BPTevol} after pre-selecting metal-rich galaxies using the criterion ${\rm N2O2} > -0.8$ (all galaxy types in a given redshift bin are displayed in a same panel in Fig. \ref{BPTevol_metalrich}). In this case, despite the low statistics of metal-rich galaxies at redshift $z>1$, SF-dominated  and active (composite and AGN-dominated) galaxies appear to be graphically separated by the optical selection criterion of \citet[][black dashed line]{Kewley01}. Yet, even when pre-selecting  metal-rich galaxies, standard optical selection criteria still fail to differentiate between composite and AGN-dominated galaxies.

\subsection{Purity and completeness fractions for optically selected galaxy types}\label{opticalfractions} 

From the previous paragraphs, we conclude that standard optical selection criteria can help better distinguish the main ionizing sources in galaxies out to high redshifts when pre-selecting metal-rich galaxies. In this subsection, we further quantify the differentiability of galaxy types with optical selection criteria over cosmic time by computing  `completeness' and `purity' fractions of SF-dominated, composite and AGN galaxies. The completeness (purity) fraction provides a measure of how complete (uncontaminated) a population of a given observationally selected galaxy type is with respect to our theoretically defined galaxy types. Specifically:
\begin{itemize}
\item The completeness fraction is computed by first selecting galaxies of a given type according to the theoretical definition, and then checking how many of these selected galaxies would be classified of the same type according to the observational selection criterion. 
\item In turn, the purity fraction is calculated by first selecting galaxies of a given type observationally, and then checking how many of these selected galaxies would be classified of the same type theoretically.
\end{itemize}

In Fig.~\ref{OpticalFrac}, we show the purity (small symbols, dashed lines in the first and third panels) and completeness  (large symbols, solid lines in the second and fourth panels)  fractions of SF-dominated (blue), composite (red) and AGN-dominated (green) galaxies versus redshift, when including all galaxies (top two panels) and only pre-selected, metal-rich galaxies  (bottom two panels). When including all galaxies, the completeness/purity fractions of SF- and AGN-dominated galaxies drop sharply from $\ga60$ per cent at low redshifts to $\la 60$ per cent at $z>1.0$, for the reasons outlined in Section~\ref{allgals}. The purity/completeness fractions of composite galaxies never exceed $\sim$40~per cent. When restricting the sample to metal-rich galaxies, the completeness/purity fractions of composite galaxies stay similarly low. In contrast, out to $z=3$ (beyond which our sample contains hardly any metal-rich galaxy), SF-dominated galaxies have completeness/purity fractions {\em always} in excess of 60/70 per cent. For AGN-dominated galaxies, the completeness fraction can even exceed 80~per cent, but the purity fraction remains low at high redshift. This is because of the difficulty of separating composite from AGN-dominated galaxies in optical line-ratio diagrams, even at high metallicity (Section~\ref{metalrich}). Still, overall, the results of  Fig.~\ref{OpticalFrac} reinforce the visual impression from Fig.~\ref{BPTevol_metalrich} that, at high interstellar metallicity, standard optical selection criteria in the \oiiihb--\niiha\ diagram help discriminate between active and inactive galaxies out to high redshift.

\section{Differentiating ionizing sources of galaxies in UV-line diagnostic diagrams}\label{UVdiagrams} 

\begin{figure*}
\textbf{UV diagnostics with 2 different emission lines incl. {\hbox{He\,II\,$\lambda1640$}}}\par\medskip\hspace{-1.5cm}
\epsfig{file=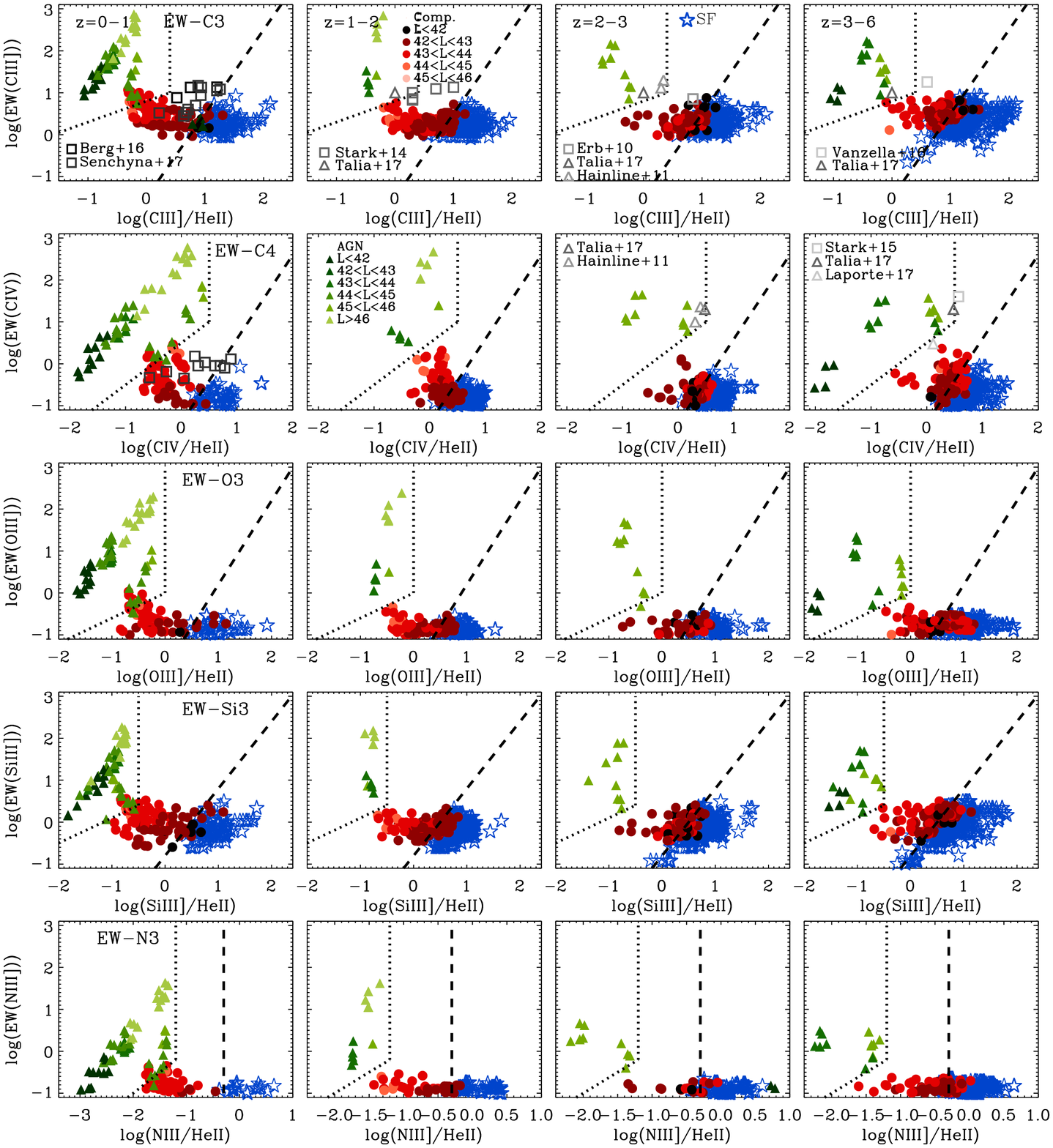,
  width=0.95\textwidth} 
\caption{Distribution of simulated metal-poor (${\rm N2O2}<-0.8$) SF (blue stars), composite (red circles) and AGN-dominated (green triangles) galaxies in five UV diagnostic diagrams constructed using  a UV line ratio and a UV line EW, in four redshift intervals, $z=0$--1, 1--2, 2--3 and 3--6 (from left to right). The diagnostic diagrams are (from top to bottom): (i) EW(\ciii) versus \ciii/\heii; (ii) EW(\civ) versus \civ/\heii; (iii) EW(\oiiiuv) versus \oiiiuv/\heii; (iv) EW(\siliii) versus \siliii/\heii; and (v) EW(\niii) versus \niii/\heii. For composite and AGN-dominated galaxies, the symbols are further colour-coded according to AGN luminosity (as indicated). As in Fig.~\ref{BPTevol}, the line ratios of simulated galaxies are shown for two different values of four undetermined parameters ($\xi_d=0.3,\,0.5$; $m_{up}=100,\,300 M_\odot$; $n_{H,\star}=10^2,\,10^3 \mathrm{cm}^{-1}$; $\alpha = -1.4,\,-2$). Also show in each diagram are the criteria from the top section of Table~\ref{Selcrit} to separate SF-dominated from composite (dashed line), and composite from AGN-dominated (dotted line) galaxies. Observations of SF-dominated \citep{Vanzella16, Berg16, Senchyna17, Stark15, Stark14, Erb10} and AGN-dominated \citep{Hainline11, Talia17, Laporte17} galaxies are reported in the first two rows, as indicated.}\label{UV2lines}    
\end{figure*}

In the previous section, we have seen that optical line-ratio diagnostic diagrams can help distinguish active from inactive galaxies out high redshifts only for metal-rich galaxies. In this section, we explore diagnostic diagrams based on {\it rest-frame UV emission lines} to discriminate between ionizing sources in metal-poor galaxies at all redshifts. As described in Section~\ref{intro}, our methodology, based on the self-consistent modelling of SF, composite and AGN-dominated galaxies in a full cosmological framework, can bring new insight over previous pioneering studies in this area \citep{Feltre16, Nakajima18}. 

In the following subsections, we start by exploring the usefulness of different combinations of UV line ratios and EW, involving (Section~\ref{UVwithHe}) or not (Section~ \ref{UVnoHe}) the \heii\ line, to classify simulated, metal-poor galaxies with different types of ionizing radiation, as defined through the theoretical BHAR/SFR ratio (see Section~\ref{opticaldiagrams}). The \heii\ line is problematic because of the difficulties of the photoionisation models in reproducing the profile and strength of this line in observed spectra (see section \ref{HeII} for discussion). These diagnostic diagrams allow us to derive selection criteria for different galaxy types (summarised in Tables~\ref{Selcrit} and \ref{Selcrit_2}). We also quantify the redshift evolution of the purity and completeness fractions for UV-selected galaxy types (Section~\ref{UVfractions}). 

In all UV diagrams considered below, we plot the same simulated galaxies in different redshift intervals as plotted in the optical diagrams of Section~\ref{opticaldiagrams}, but including now only metal-poor galaxies, as pre-selected with ${\rm N2O2}< -0.8$ (i.e. with metallicities below about half solar; see Section~\ref{metalrich}). Since we focus on such metal-poor galaxies, we neglect any contribution from post-AGB stellar populations to the total nebular emission \citep[this contribution is mostly important for evolved, metal-rich galaxies at low redshift; see e.g.][]{Hirschmann17}. By analogy with the case of optical lines in Section~\ref{opticaldiagrams}, we apply a flux detection limit of $\rm 10^{-18}\,erg\,s^{-1}cm^{-2}$ to all UV emission lines of simulated galaxies (note that we typically lose less then 10~per cent of our galaxies, when applying this flux detection limit). We also apply a detection limit of 0.1~\AA\ on UV emission-line EW.\

\subsection{UV-line diagnostics of \textit{metal-poor} galaxies including {\hbox{He\,II\,$\lambda1640$}}}\label{UVwithHe} 

\begin{figure*}
\center
\textbf{UV diagnostics with 3 different emission lines incl. {\hbox{He\,II\,$\lambda1640$}}}\par\medskip\hspace{-1.5cm}
\epsfig{file=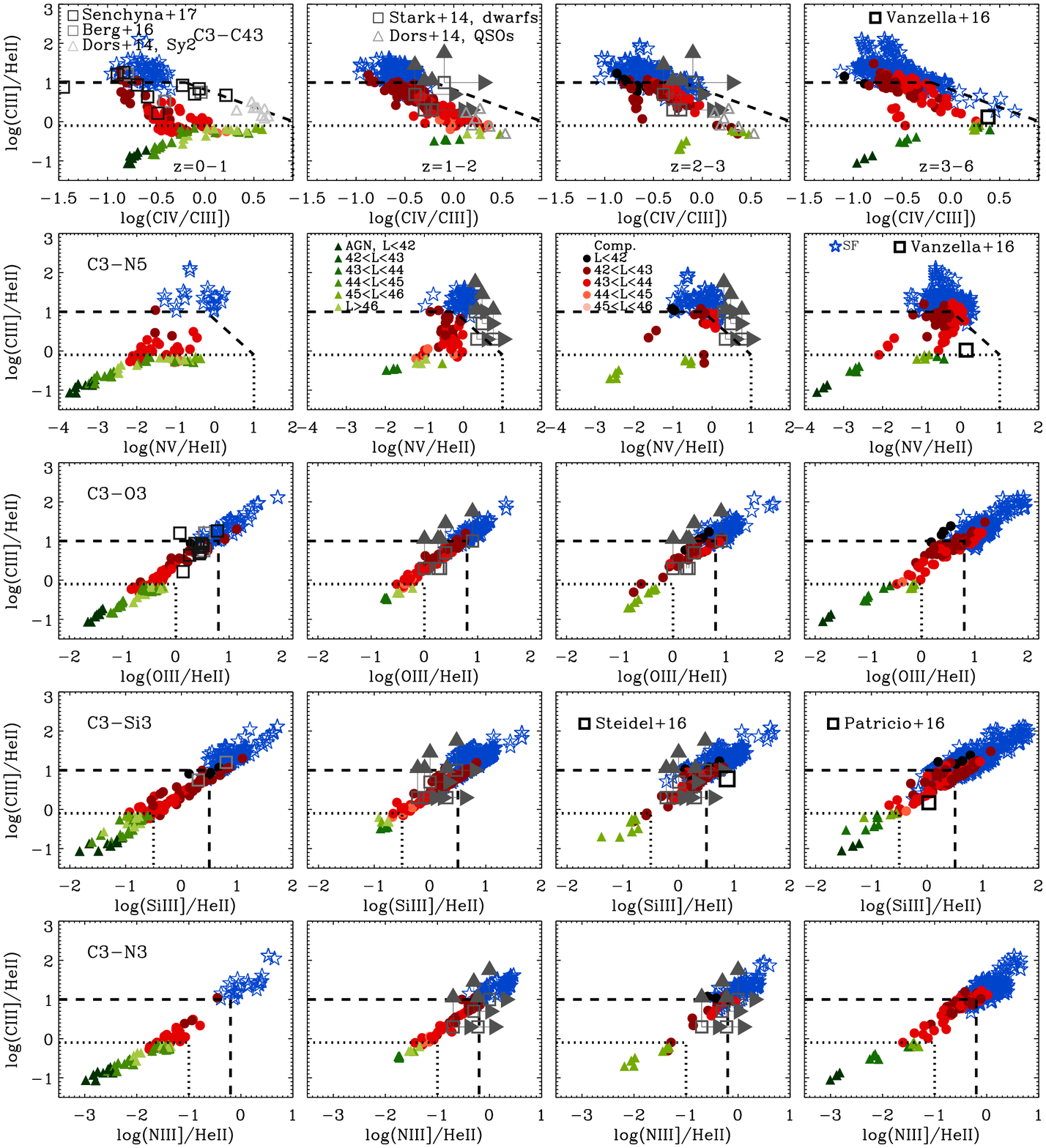,
  width=0.95\textwidth} 
\caption{Same as Fig. \ref{UV2lines}, but for UV diagnostic diagrams constructed using line ratios only. The diagrams are (from top to bottom):  (i) \ciii/\heii\ versus \ciii/\civ; (ii) \ciii/\heii\ versus \nv/\heii; (iii) \ciii/\heii\ versus \oiiiuv/\heii; (iv) \ciii/\heii\ versus \siliii/\heii; and (v) \ciii/\heii\ versus \niii/\heii. Also show in each diagram are the criteria from the middle section of Table~\ref{Selcrit} to separate SF-dominated from composite (dashed line), and composite from AGN-dominated (dotted line) galaxies. Observations of dwarf galaxies at $z\sim0$ \citep{Berg16, Senchyna17} and $z=2$--3 \citep{Stark14},  individual distant SF galaxies \citep{Patricio16,Steidel16,Vanzella16}, Seyfert-2 galaxies at $z \sim 0$ and type-2 quasars at $z\sim 2$ \citep{Dors14} are reported, as indicated.}\label{UV3lines}    
\end{figure*}

Fig.~\ref{UV2lines} shows the distribution of SF (blue stars), composite (red circles) and AGN-dominated (green triangles) galaxies in five UV diagnostic diagrams constructed using  a UV line ratio and a UV line EW, in four redshift intervals, $z=0$--1, 1--2, 2--3 and 3--6 (from left to right). The diagnostic diagrams are (from top to bottom):\\\vspace{-0.2cm}

\noindent (i) EW(\ciii)) versus \ciii/\heii\ (hereafter EW-C3);\\ 
(ii) EW(\civ)) versus \civ/\heii\ (hereafter EW-C4);\\ 
(iii) EW(\oiiiuv) versus \oiiiuv/\heii\ (hereafter EW-O3);\\ 
(iv) EW(\siliii) versus \siliii/\heii\ (hereafter EW-Si3); and\\ 
(v) EW(\niii) versus \niii/\heii\ (hereafter EW-N3).\\\vspace{-0.2cm}

For composite and AGN-dominated galaxies, the symbols are further colour-coded according to AGN luminosity (as indicated). 

For each UV diagnostic diagram shown in each redshift range, we find a fairly clear separability of different galaxy types, with only some minor overlap between SF-dominated galaxies and composite galaxies with faint AGN.  This clear differentiability of galaxy types arises primarily from differences in line ratios, but also in line EW, as most AGN-dominated galaxies have EW larger than reached by SF-dominated (and to a large extent also by composite) galaxies. The ratio of any of the five (collisionally excited) metal lines considered in Fig.~\ref{UV2lines} to the \heii\ (recombination) line decreases from SF-dominated, to composite, to AGN-dominated galaxies. This is because of the harder ionizing radiation of accreting BHs compared to stellar populations, which increases the probability of doubly ionizing helium. This hard radiation also makes the EW of high-ionization, collisionally excited metal lines stronger, and more so for luminous than for faint AGN. We recall that the EW in Fig.~\ref{UV2lines}, which account for the emission from narrow-line regions but not for direct attenuated radiation from accreting BHs, should be appropriate for type-2 AGN and taken as upper limits for type-1 AGN (see Section~\ref{eqwidth}).

Also shown in each panel of Fig.~\ref{UV2lines} are proposed observational selection criteria to separate SF-dominated from composite (dashed line), and composite from AGN-dominated (dotted line) galaxies. These selection criteria, reported in the top section of Table~\ref{Selcrit},  were chosen to maximise the purity and completeness fractions (Section~ \ref{UVfractions}). Interestingly, the same criteria remain valid over the whole redshift range from $z=0$ to 6. This indicates that the UV-line ratios and EW of metal-poor galaxies considered here depend far less sensitively on ISM properties than on the nature of the ionizing radiation.


The EW-C3 and EW-C4 diagrams (first two rows) of Fig.~\ref{UV2lines} also show available observations of dwarf SF galaxies at $z\sim 0$ \citep{Berg16, Senchyna17} and $z=1.5$--4 \citep{Stark14}, and three SF-dominated galaxies at $z =2.3$, 3.1 and 7 \citep[][respectively]{Erb10, Stark15,Vanzella16}, along with AGN-dominated galaxies at $z=2$--3, 1.4--4.6 and 7 \citep[respectively]{Hainline11, Talia17, Laporte17}, as indicated. The predictions from our simulations are in fair agreement with these sparse observational data: observed SF- and AGN-dominated galaxies nicely fall in the regions corresponding to our theoretically defined SF/composite and AGN categories. This agreement with observations corroborates our theoretically derived selection criteria. We note that very high-quality observations will be required to fully exploit the selection criteria of Fig.~\ref{UV2lines}, which require EW measurements down to 0.1\,\AA.

\begin{figure*}
\textbf{UV diagnostics with 3 or 4 different emission lines without {\hbox{He\,II\,$\lambda1640$}}}\par\medskip\hspace{-1.5cm}
\epsfig{file=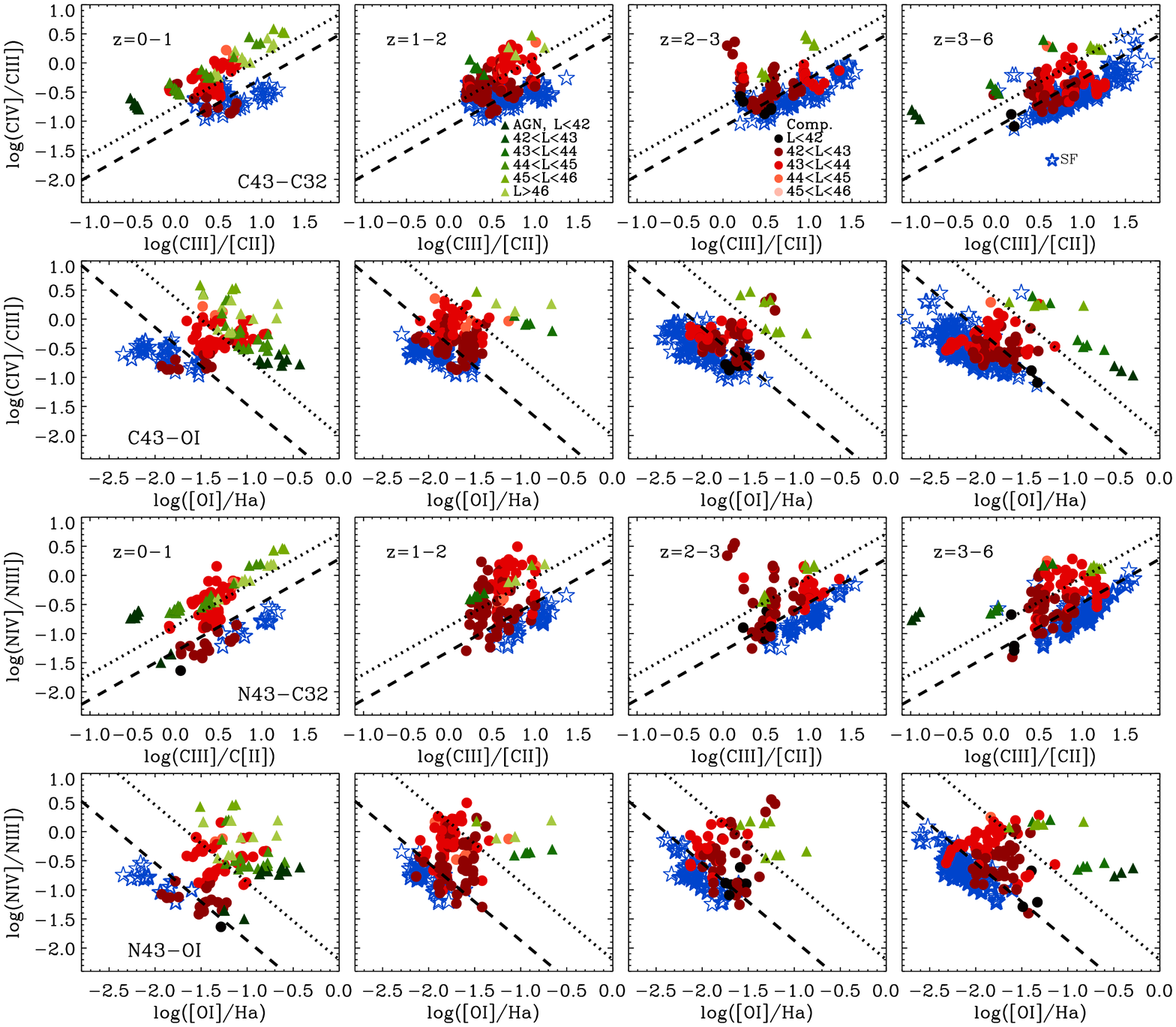,
  width=0.95\textwidth} 
\caption{Same as Fig.~\ref{UV2lines}, but for combined UV/optical diagnostic diagrams not including the \heii\ line. The diagrams are (from top to bottom): (i) \civ/\ciii\ versus \ciii/\cii; (ii) \civ/\ciii\ versus \oi/\ha; (iii) \niv/\niii\ versus \ciii/\cii; and (iv) \niv/\niii\ versus \oi/\ha. Also show in each diagram are the criteria from the bottom section of Table~\ref{Selcrit} to separate SF-dominated from composite (dashed line), and composite from AGN-dominated (dotted line) galaxies.}\label{UV4lines}    
\end{figure*}

\begin{table*}
\centering
\begin{tabular}{ p{3cm} | p{1.5cm} | p{3.5cm} | p{3.5cm} | p{3.5cm} p{0.001cm}}
\centering {\bf UV diagram} & \centering {\bf Acronym} & \centering {\bf SF galaxies}  & \centering {\bf Composites}  & \centering {\bf AGN} & \\ \hline \hline
 \multicolumn{6}{c}{\bf Criteria based on a UV line ratio and a UV line EW, including {\hbox{He\,II\,$\lambda1640$}}} \\ \hline \hline
 
 y = log(EW(\ciii)), & \centering {\bf EW-C3} & \centering $y < 2x-1.5$ & \centering $y > 2x-1.5$, & \centering $y > 0.5x+0.8$, $x<0.4$ &  \\ 
  x = log(\ciii/\heii) & & & \centering $y < 0.5x+0.8$ if $y < 1$, & &  \\ 
 & & & \centering $x > 0.4$ if $y>1$ & & \\ \hline
 
 y = log(EW(\civ)), & \centering {\bf EW-C4} & \centering $y <  2x-1.4$ & \centering $y >  2x-1.4$ & \centering $y > x+0.5$, $x<0.5$ & \\
 x = log(\ciii/\heii) & &  & \centering $y < x+0.5$ if $y < 1$ & & \\
  & &  & \centering $x>0$ if $y > 1$ & & \\ \hline
  
  y = log(EW(\oiiiuv)), & \centering {\bf EW-O3} & \centering $y <  2x-1.8$ & \centering $y >  2x-1.8$ & \centering $y > 0.6x$, $x<0$ & \\
 x = log(\oiiiuv/\heii) & &  & \centering $y < 0.6x$ if $y < 0$ & & \\
  & &  & \centering $x>0$ if $y > 0$ & & \\ \hline
  
   y = log(EW(\siliii)), & \centering {\bf EW-Si3} & \centering $y <  1.6x-0.8$ & \centering $y >  1.6x-0.8$ & \centering $y > 0.6x+0.7$, $x<-0.5$ & \\
 x = log(\siliii/\heii) & &  & \centering $y < 0.6x+0.7$ if $y < 0.3$ & & \\
  & &  & \centering $x>-0.5$ if $y > 0.3$ & & \\ \hline
  
    y = log(EW(\niii)), & \centering {\bf EW-N3} & \centering $x > -0.3$ & \centering $x < -0.3$ & \centering $y > x+1$, $x<-1.2$ & \\
 x = log(\niii/\heii) & &  & \centering $y < x+1$ if $y < -0.3$ & & \\
  & &  & \centering $x>-1.2$ if $y > -0.3$ & & \\ \hline \hline
  
\multicolumn{6}{c}{\bf Criteria based on ratios of three UV lines, including {\hbox{He\,II\,$\lambda1640$}}} \\ \hline \hline
   
   y = log(\ciii/\heii), & \centering {\bf C3-C43} & \centering $y > -0.9x+0.8$ if $x>-0.2$, & \centering $y < -0.9x+0.8$ if $x>-0.2$, & \centering $y < -0.1$ &  \\ 
  x = log(\civ/\ciii) & & \centering $y>1$ if $x<-0.2$ & \centering $y<1$ if $x<-0.2$, & &  \\ 
 & & & \centering $y>-0.1$ & & \\ \hline
 
  y = log(\ciii/\heii), & \centering {\bf C3-N5} & \centering $y > -0.9x+0.8$ if $x>-0.2$, & \centering $y < -0.9x+0.8$ if $x>-0.2$, & \centering $y < -0.1$, $x<1$ &  \\ 
  x = log(\nv/\heii]) & & \centering $y>1$ if $x<-0.2$ & \centering $y<1$ if $x<-0.2$, & &  \\ 
 & & & \centering $y>-0.1$ & & \\ \hline
 
  y = log(\ciii/\heii), & \centering {\bf C3-O3} & \centering $y > 1$ or $x>0.8$ & \centering $y < 1$, $x<0.8$, & \centering $y < -0.1$, $x<0$ &  \\ 
  x = log(\oiiiuv/\heii]) & &  & \centering $y>-0.1$ or $x>0$ & &  \\ \hline
  
y = log(\ciii/\heii), & \centering {\bf C3-Si3} & \centering $y > 1$ or $x>0.5$ & \centering $y < 1$, $x<0.5$, & \centering $y < -0.1$, $x<-0.5$ &  \\ 
x = log(\siliii/\heii]) & &  & \centering $y>-0.1$ or $x>-0.5$ & &  \\ \hline
  
  y = log(\ciii/\heii), & \centering {\bf C3-N3} & \centering $y > 1$ or $x>-0.2$ & \centering $y < 1$, $x<-0.2$, & \centering $y < -0.1$, $x<-1$ &  \\ 
  x = log(\niii/\heii]) & &  & \centering $y>-0.1$ or $x>-1$ & &  \\ \hline

\end{tabular}
\caption{Observational selection criteria to discriminate graphically between metal-poor (${\rm N2O2}<-0.8$) SF-dominated, composite and AGN-dominated galaxies, derived from distributions of simulated galaxies with different BHAR/SFR ratios in the 14 UV/optical diagnostic diagrams of Figs~\ref{UV2lines}, \ref{UV3lines} and \ref{UV4lines}. The first column specifies the abscissa and ordinate (EW or line ratio) and the second column the acronym of the diagnostic diagram; the third, fourth and fifth columns the individual selection criteria for SF-dominated, composite and AGN-dominated galaxies, illustrated by the dashed (SF-composite) and dotted (composite-AGN) lines in Figs~\ref{UV2lines} and \ref{UV3lines}. See text in Sections~\ref{UVwithHe} and \ref{UVnoHe} for details.}\label{Selcrit}
\end{table*}

\begin{table*}
\centering
\begin{tabular}{ p{3cm} | p{1.5cm} | p{3.5cm} | p{3.5cm} | p{3.5cm} p{0.001cm}}
\centering {\bf UV diagram} & \centering {\bf Acronym} & \centering {\bf SF galaxies}  & \centering {\bf Composites}  & \centering {\bf AGN} & \\ \hline \hline

\multicolumn{6}{c}{\bf Criteria based on ratios of UV and optical lines, not including {\hbox{He\,II\,$\lambda1640$}}} \\ \hline \hline

 y = log(\civ/\ciii), & \centering {\bf C43-C32} & \centering $y < 0.8x-1.1$ & \centering $y > 0.8x-1.1$, & \centering $y > 0.8x-0.75$ &  \\ 
  x = log(\ciii/\cii) & &  & \centering $y < 0.8x-0.75$ & &  \\ \hline
  
  y = log(\civ/\ciii), & \centering {\bf C43-OI} & \centering $y < -1.3x-3.2$ & \centering $y > -1.3x-2.8$, & \centering $y > -1.3x-2$ &  \\ 
  x = log(\oi/H$\alpha$) & &  & \centering $y < -1.3x-2$ & &  \\ \hline   

  y = log(\niv/\niii), & \centering {\bf N43-C32} & \centering $y < 0.8x-1.3$ & \centering $y > 0.8x-1.3$, & \centering $y > 0.8x-0.9$ &  \\ 
  x = log(\ciii/\cii) & &  & \centering $y < 0.8x-0.9$ & &  \\ \hline
  
  y = log(\niv/\niii), & \centering {\bf N43-OI} & \centering $y < -1.3x-3.2$ & \centering $y > -1.3x-3.2$, & \centering $y > -1.3x-2.2$ &  \\ 
  x = log(\oi/H$\alpha$) & &  & \centering $y < -1.3x-2.2$ & &  \\ \hline

\end{tabular}
\caption{Same as in Table 1, but now for diagnostic diagrams not including the \heii\ line (see selection criteria quantified by the dashed (SF-composite) and dotted (composite-AGN) lines in Fig. \ref{UV4lines}).}\label{Selcrit_2}
\end{table*}

 In Fig~\ref{UV3lines}, we show an alternative to Fig.~\ref{UV2lines} constructed using line ratios only. Specifically, we plot the distribution of SF, composite and AGN-dominated galaxies in five UV diagnostic diagrams constructed each with the \ciii/\heii\ ratio in ordinate and a ratio involving a third line in abscissa, at different redshifts. The diagrams are (from top to bottom): \\\vspace{-0.2cm}

\noindent (i) \ciii/\heii\ versus \ciii/\civ\ (hereafter C3-C43); \\
\noindent (ii) \ciii/\heii\ versus \nv/\heii\ (hereafter C3-N5); \\
\noindent (iii) \ciii/\heii\ versus \oiiiuv/\heii\ (hereafter C3-O3); \\
\noindent (iv) \ciii/\heii\ versus \siliii/\heii\ (hereafter C3-Si3); and \\
\noindent (v) \ciii/\heii\ versus \niii/\heii\ (hereafter C3-N3).\\\vspace{-0.2cm}

Irrespective of the redshift range, we obtain a fairly clear separability of different types of metal-poor galaxies in all UV diagnostic diagrams shown. 

The clear separability of different galaxy types in Fig~\ref{UV3lines} arises from the fact that ratios of collisionally-excited metal lines to the \heii\ recombination line are highly sensitive to the hard radiation from an AGN, as described for Fig.~\ref{UV2lines} above. We note that this is less the case for \nv/\heii\ (second row), since \nv\ requires photons of even higher energy than \heii\ to be produced (77.5 versus 54.4\,eV). Hence, \nv/\heii\ is not a clear indicator of the hardness of the ionizing radiation. In the C3-C43 diagnostic diagram (top row), the \civ/\ciii\ ratio is also sensitive to the rise of energetic photons from the presence of an AGN, which increases the probability of triply ionizing carbon. 

As in Fig.~\ref{UV2lines}, also shown in each panel of Fig.~\ref{UV3lines} are proposed observational selection criteria to separate SF-dominated from composite (dashed line), and composite from AGN-dominated (dotted line) metal-poor galaxies. These selection criteria, reported in bottom section of Table~\ref{Selcrit}, can be applied over the whole redshift range from $z=0$ to 6. 

To compare our theoretical predictions with observations of UV emission lines in active and inactive galaxies, we show, when available, observations from three main samples in the different panels of Fig.~\ref{UV3lines}, as indicated: (i) a sample of 22 type-2 AGN assembled by \citet[][and references therein]{Dors14}, consisting of 12 Seyfert-2 galaxies in the local Universe and 10 X-ray selected type-2 quasars at redshift $1.5 < z < 4.0$; (ii) a sample of five gravitationally lensed, low-mass star-forming galaxies at redshifts $1.5 < z < 3.0$ from \citet{Stark14}; and (iii) two samples of dwarf galaxies at $z\sim 0$ from \citet[][]{Senchyna17} and \citet{Berg16}. In addition, we show a few individual distant galaxies from \citet{Vanzella16}, \citet{Patricio16} and \citet{Steidel16}.

We find that observations of SF-dominated galaxies in Fig.~\ref{UV3lines} generally overlap with the theoretically defined SF-galaxy regions (above the dashed line in each panel), but can also fall in the composite-galaxy regions (between dashed and dotted lines). This does not necessarily indicate a mismatch between predictions and observations, as observed SF-dominated galaxies may include a minor contribution from a central accreting BH (or the \heii\ line is not properly modelled, see section \ref{HeII} for further discussion). Also, some dwarf-galaxy measurements provide only limits on emission-line ratios \citep{Stark14}, compatible with both composite and SF-dominated model galaxies. Similarly, observations of AGN-dominated galaxies generally overlap with the theoretically defined AGN regions in Fig.~\ref{UV3lines}, but can also fall in the composite-galaxy regions. This could arise from the contamination of line-flux measurements of AGN narrow-line regions by star formation in the host galaxy \citep[e.g.,][]{Bonzini13, Antonucci15}. We conclude that, overall, the sparse observational data currently available are compatible with the UV selection criteria in Fig.~\ref{UV3lines}.

We note that the line measurements shown in Figs.~\ref{UV2lines} and \ref{UV3lines} were not corrected for potential attenuation by dust in the galaxies. This is not important for the samples of low-mass star-forming galaxies, which have been shown to be extremely dust-poor (e.g., table~7 of \citealp{Stark14}). For the AGN samples, we prefer not to correct the observed fluxes using an arbitrary attenuation curve, as the dispersion between different standard curves is large at UV wavelengths (e.g., figure 9 of \citealp{Charlot00}). Instead, when comparing synthetic UV lines with these data, we computed the attenuation that would be inferred using the \citet{Calzetti00} curve for a $V$-band attenuation of one magnitude ($A_V = 1$) and found the impact on our analysis to be largely negligible  (see Section~\ref{dust} for a more detailed discussion).

\subsection{UV-line diagnostics of \textit{metal-poor} galaxies not including {\hbox{He\,II\,$\lambda1640$}}}\label{UVnoHe} 

All the UV-diagnostic diagrams we considered so far to discriminate between ionizing sources in metal-poor galaxies included the \heii\ line. Recently, problems have arisen when fitting this emission line in observed spectra of local dwarf and high-redshift SF galaxies, using state-of-the-art photo-ionization models combined with (single-star or binary)  stellar population synthesis models \citep[e.g.,][]{Gutkin16, Steidel16}: in some cases, particularly for very metal-poor dwarf galaxies with metallicity below 1/5 solar, models can underestimate the observed \heii-line luminosity by almost an order of magnitude \citep[e.g.,][]{Senchyna17, Steidel16, Berg18, Jaskot16}. In this context, the reliability of the UV selection criteria presented in Section~\ref{UVwithHe} to discriminate between SF-dominated, composite and AGN-dominated metal-poor galaxies may be reduced (see Section~\ref{HeII} for a more detailed discussion). To circumvent this potential drawback, we now present two alternative UV and two combined optical-UV diagnostic diagrams to discriminate between different galaxy types, {\em which avoid the problematic \heii\ line}. 

By analogy with Figs~\ref{UV2lines} and \ref{UV3lines}, we show in Fig.~\ref{UV4lines} the distribution of SF, composite and AGN-dominated galaxies in four different diagnostic diagrams at different redshifts. The diagrams are (from top to bottom):\\\vspace{-0.2cm} 

\noindent (i) \civ/\ciii\ versus \ciii/\cii\ (hereafter C43-C32);\\
(ii) \civ/\ciii\ versus \oi/\ha\ (hereafter C43-O1);\\
(iii) \niv/\niii\ versus \ciii/\cii\ (hereafter N43-C32); and\\ 
(iv) \niv/\niii\ versus \oi/\ha\ (hereafter N43-O1).\\\vspace{-0.2cm} 

In all diagrams, different galaxy types fall into different regions, but generally the separability appears to be less clear in these new diagnostic diagrams without the \heii\ line than in the previous ones including the \heii\ line (see Figs. \ref{UV2lines} and \ref{UV3lines}). 

The \civ/\ciii\ and \niv/\niii\ ratios typically increase from SF-dominated, to composite, to AGN-dominated galaxies and with AGN luminosity, because of the corresponding larger fraction of highly energetic photons (Section~\ref{UVwithHe}). The \ciii/\cii\ ratio is less sensitive to the details of the ionizing spectrum at high energies, while the \oi/\ha\  ratio is slightly larger for composite and AGN-dominated galaxies than for SF-dominated galaxies, for reasons outlined in \citet{Hirschmann17}. A drawback of the combined UV-optical diagnostic diagrams in Fig.~\ref{UV4lines} is that optical lines of distant galaxies shift beyond the range of near-infrared spectrographs at $z \ga 6$. We note that replacing \oi/\ha\ by \sii/\ha\ in Fig.~\ref{UV4lines} would result in a similar separability between different galaxy types. We also found that many other UV line-ratio combinations without the \heii\ line, for example \siliv/\siliii\ and \siliii/\silii, cannot help discriminate between ionizing sources in galaxies.

As in Figs~\ref{UV2lines} and \ref{UV3lines}, we propose in each panel of Fig.~\ref{UV4lines} two observational selection criteria to separate SF-dominated from composite (dashed line), and composite from AGN-dominated (dotted line) metal-poor galaxies. These selection criteria are reported in Table~\ref{Selcrit_2}. We could not find any readily available observations in the literature to test/validate these criteria.

\subsection{Purity and completeness fractions for UV-selected galaxy types}\label{UVfractions}

\begin{figure*}
\epsfig{file=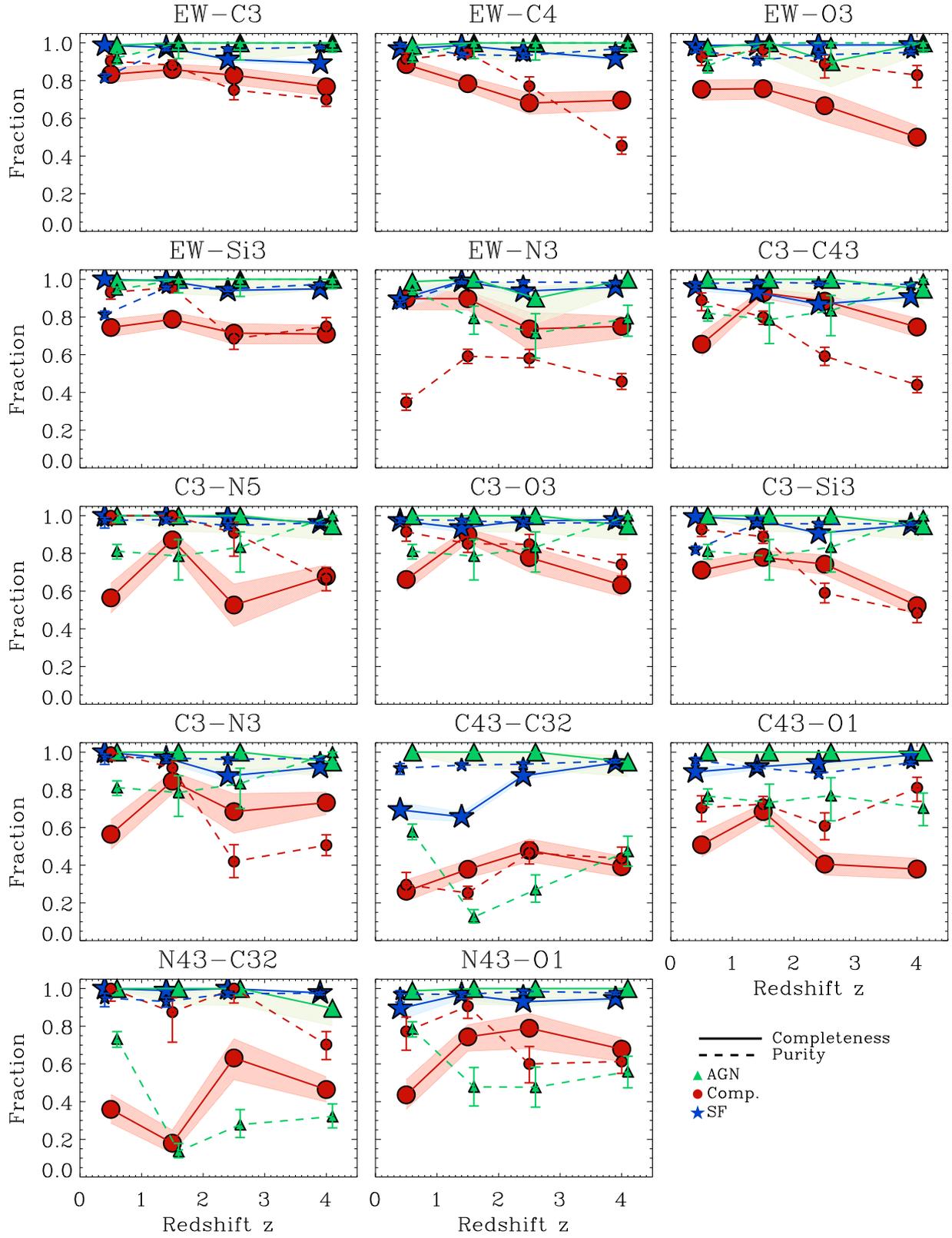,
  width=0.95\textwidth} 
\caption{Purity (small symbols, dashed lines) and completeness (large symbols, solid lines) fractions as a function of redshift, for simulated metal-poor (${\rm N2O2}<-0.8$) SF-dominated (blue stars and lines), composite (red circles and lines) and AGN-dominated (green triangles and lines) galaxies, as classified observationally using the UV selection criteria of Table~\ref{Selcrit}. Each panel corresponds to a different UV diagnostic diagram from Figs~\ref{UV2lines}, \ref{UV3lines} and \ref{UV4lines}, as indicated at the top.}\label{UVFrac}      
\end{figure*}

In the previous two subsections, we identified 14 UV diagnostic diagrams to graphically distinguish between different galaxy types via observational selection criteria reported in Tables~\ref{Selcrit} and \ref{Selcrit_2}. In this section, we investigate the reliability of these UV selection criteria by computing the corresponding completeness and purity fractions, as defined in Section~\ref{opticalfractions}.

Fig.~\ref{UVFrac} shows the completeness (large symbols, solid lines) and purity (small symbols, dashed lines) fractions of SF-dominated (blue), composite (red) and AGN-dominated (green) galaxies versus redshift, as selected using the criteria in Figs~\ref{UV2lines}--\ref{UV4lines}. The 14 panels correspond to the 14 different diagnostic diagrams of Figs~\ref{UV2lines}--\ref{UV4lines}. In all cases, the purity/completeness fractions of SF-dominated galaxies and the completeness fraction of AGN-dominated galaxies stay mostly above 90~per cent over the entire redshift range (except for C43-C32). This indicates that our proposed UV selection criteria work remarkably well to select SF-dominated galaxies, and complete samples of AGN-dominated galaxies, from the simulation galaxy sample.

For the EW-C3, EW-C4, EW-O3, EW-Si3 diagrams, the selected AGN samples are in addition largely uncontaminated by other galaxy types (in particular, composites), as indicated by AGN purity fractions in excess of 90~per cent. The EW-N3, C3-C43, C3-N5, C3-O3, C3-Si3, and C3-N3 diagrams also provide fairly high AGN purity fractions, above 75~per cent. In contrast, for the C43-C32, C43-OI, N43-C32 and N32-O1 diagrams not including the \heii\ line, AGN purity fractions can drop fairly low, down to only 20~per cent, as a consequence of contamination by composite galaxies.

As in the case of optical selection criteria (Section~\ref{opticalfractions}), the UV-derived purity/completeness fractions of {\em composite} galaxies fall generally below those of SF-  and AGN-dominated galaxies. In particular, diagnostic diagrams not including the \heii\ line are characterised by low purity and/or completeness fractions of composites (down to 20~per cent), as composites often overlap with SF- and AGN-dominated regions in these diagrams. The highest purity/completeness fractions for composites are reached with the EW-C3 ($>75$~per cent), as well as the EW-Si3 and C3-O3 ($>60$~per cent) diagrams. Thus, we identify these UV diagnostic diagrams as most powerful to reliably discriminate between ionizing sources in not only local, but also distant metal-poor galaxies. As mentioned before, this may require accurate measurements of faint line EW.

\section{Discussion}\label{discussion} 

In the last two sections, our combination of zoom-in galaxy simulations with versatile nebular-emission models allowed us to show that optical `BPT' diagnostic diagrams can help  discriminate between ionizing sources in metal-rich galaxies out to high redshifts, and UV diagnostic diagrams in {\em metal-poor} galaxies at all redshifts. We quantified novel selection criteria based on 14 UV diagnostic diagrams (reported in Tables~\ref{Selcrit} and \ref{Selcrit_2}), and found the EW-C3, EW-Si3 and C3-O3 line-ratio/EW combinations to be most sensitive to the nature of the ionizing source. In this section, we discuss some caveats of our derived UV selection criteria (Sections~\ref{dust}, \ref{HeII} and \ref{shocks}) and reframe our results on UV selection criteria into the context of previous theoretical studies (Section~\ref{comp}).


\begin{figure*}
\epsfig{file=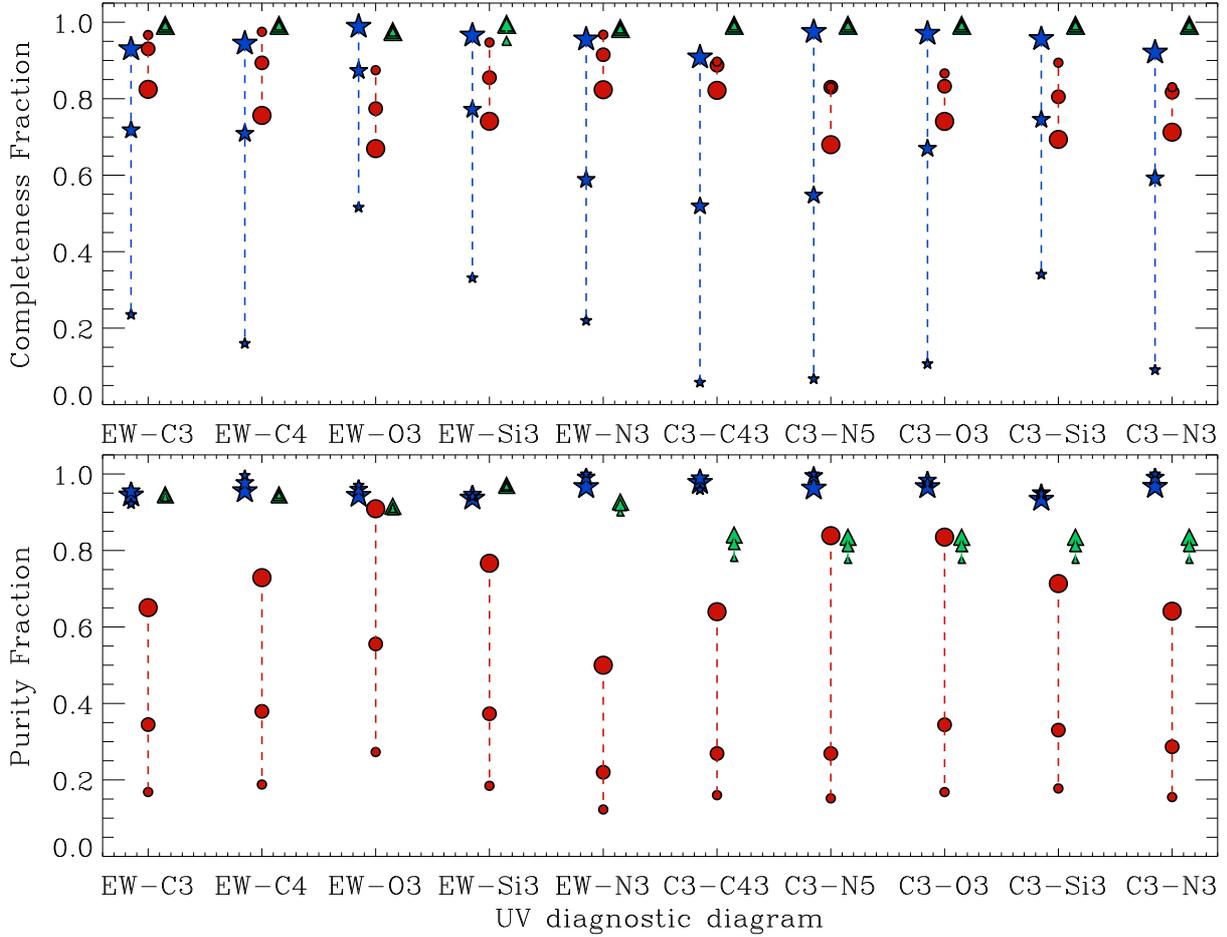,
  width=0.95\textwidth} 
\caption{Purity (bottom panel) and completeness (top panel) fractions for simulated metal-poor (${\rm N2O2}<-0.8$) SF-dominated (blue stars and lines), composite (red circles and lines) and AGN-dominated (green triangles and lines) galaxies, as classified observationally using the UV selection criteria of Tables~\ref{Selcrit} and \ref{Selcrit_2}. Each abscissa corresponds to a different UV diagnostic diagram from Figs~\ref{UV2lines} and \ref{UV3lines}, as indicated. Large symbols correspond to the same models as in Fig.~\ref{UVFrac}. Medium symbols correspond to models in which the \heii-line luminosity from young stars was artificially increased by a factor of 2, and small symbols to models in which the \heii-line luminosity was artificially increased by a factor of 4.}\label{HeIIfrac}      
\end{figure*}

\subsection{Dust attenuation}\label{dust}

 As mentioned in Section~\ref{UVwithHe}, to explore the effect of dust attenuation by the diffuse interstellar medium on synthetic UV lines of a galaxy,\footnote{Instead, within one HII region or narrow line region, dust attenuation is self-consistently accounted for by \textsc{Cloudy}.} we computed the effect of attenuated emission lines that would be inferred using the \citet{Calzetti00} curve for a $V$-band attenuation of one magnitude ($A_V = 1$). We found the impact on our analysis to be largely negligible. Specifically, line ratios, such as \ciii/\heii, \civ/\heii, \oiiiuv/\heii, \siliii/\heii, \niii/\heii\ and \civ/\ciii\ would be shifted by less than $\sim0.1$\,dex by such dust attenuation; only \niv/\heii\ would be lower by about $\sim$0.3 dex \citep[see also][]{Feltre16}. The effect on diagnostic diagrams is modest, therefore, when adopting the same attenuation prescription for all lines. However, the effect could be larger if, for example, the regions producing different emission lines where attenuated in different ways. As an example, this differential dust attenuation might become important for the \niv/\heii-ratio, as the \niv\ is mostly produced in the NLR, while \heii\ can be produced in both NLR and HII regions.


In order to minimize the impact of dust attenuation, optical emission lines used in line ratios of the classical BPT diagrams were chosen to be as close as possible to one another in wavelength ($\Delta \lambda$). Their differences in wavelength vary between $\Delta \lambda = 20$~\AA\ (for \niiha) and $\Delta \lambda = 260$~\AA\ (for \oiha). Most of the line ratios in our suggested UV diagnostic diagrams involve lines with a wavelength difference smaller than 250~\AA, except for \civ/\ciii\ ($\Delta \lambda = 358$~\AA), \nv/\heii\ ($\Delta \lambda = 400$~\AA) and \ciii/\cii\ ($\Delta \lambda = 418$~\AA). Thus, the UV diagnostic diagrams C3-C43, C3-N5, N43-C32 and C43-C32 might be affected more strongly by dust than the other UV diagrams proposed in this study. However, as the attenuation by dust is stronger in the UV than in the optical, at fixed line wavelength difference, its impact can be stronger in UV than in optical diagrams even for $\Delta \lambda < 250$~\AA. A sophisticated modelling of dust will be certainly needed to robustly address the effect of dust on UV diagnostic diagrams, which, is, however, beyond the scope of this work.

\subsection{The problematic {\hbox{He\,II\,$\lambda1640$}}\ line in photoionization models}\label{HeII}

As mentioned in Section~\ref{UVnoHe} above, all currently available models of nebular emission appear to severly underestimate the observed \heii-line luminosity in some extreme cases of very metal-poor star-forming galaxies \citep[with $Z_{\rm gas} \la 0.2\,Z_\odot$; e.g.,][]{Jaskot16,Steidel16,Senchyna17,Berg18}. The potential reasons for this discrepancy between models and observations are still heavily debated in the literature.

At least for the dwarf galaxies studied by \citet{Senchyna17}, extra photoionization by an AGN is unlikely to ease the tension between observed and modelled \heii\ luminosities. This study also showed that for some observed spectra, the models of \citet{Gutkin16} can fit the \heii\ line when invoking an IMF extending out to 300 $M_\odot$, but only with a metallicity substantially lower than the optically derived gas-phase metallicity, hence not providing any satisfactory solution. In addition, photoionization models including binary stellar population models (e.g., BPASS, \citealp{Eldridge17}) appear to also fail to reproduce observed \heii\ lines of some metal-poor star-forming galaxies \citep{Shirazi12, Senchyna17}.

This leaves three primary ionizing sources, which could contribute to enhanced \heii\ emission: stars, shock excitation and X-ray binaries \citep[see section 5.3 of][for a detailed discussion]{Senchyna17}. New observations of extremely metal-poor galaxies by \citet{Senchyna18} show that unlike for the other nebular lines, the equivalent width of \heii\ does not scale with that of \hb\ in their sample. This would suggest that  some extremely high-energy ($> 54.4$~eV)  flux in these very metal-poor galaxies is produced by sources with timescales $>$ 10 Myr (effectively decoupled from the lifetimes of most massive stars), such as stars stripped by close binary interactions or low-mass X-ray binaries, and shocks. Such processes are not yet included in nebular emission models.

The uncertainties affecting models of the \heii\ line in metal-poor galaxies cast some doubt about the validity of the UV selection criteria presented in Section~\ref{UVwithHe}. To assess the robustness of these criteria, we can study the impact of {\em artificially} increasing the \heii-line luminosity predicted by our models. Fig.~\ref{HeIIfrac} illustrates the completeness (top panel) and purity (bottom panel) fractions of SF-dominated, composite and AGN-dominated (different symbols) metal-poor galaxies at redshifts $z=0$--6, for the 10 UV diagnostic diagrams including the \heii\ line (as indicated on the x-axis), after multiplying the predicted \heii-line luminosities from young stars by factors of 1 (i.e., fiducial model, large symbols), 2 (middle-size symbols) and 4 (small symbols). 

Remarkably, Fig.~\ref{HeIIfrac} shows that the purity/completeness fractions of AGN-dominated galaxies and the purity fractions of SF-dominated galaxies are {\em hardly} affected by an increased \heii\ luminosity and remain above 80~per cent. Also the completeness fractions of composites show comparably little variation with increasing \heii\ luminosity, staying largely above 70~per cent. This indicates that UV-selected AGN-dominated galaxies can still be reliably identified, SF-dominated samples are widely uncontaminated by active galaxies (composites and AGN), and composite samples are fairly complete. 

In contrast, the completeness fractions of SF-dominated galaxies and the purity fractions of composites can drop below 20~per cent if the \heii\ luminosity increases by a factor of four. This is because a rise in the \heii\ luminosity of an SF-dominated galaxy makes the ratio of any metal line to \heii\ move toward the region populated by composites in UV diagnostic diagrams. As a consequence, SF-dominated samples are not complete anymore, and composite samples are strongly contaminated by SF-dominated galaxies. We note that modifying the UV selection criteria does not provide any acceptable solution because of the intrinsic overlap of SF-dominated and composite galaxies in ratios of a metal line to \heii. 

Fig. \ref{HeIIfrac} further indicates that the purity/completeness fractions in the EW-O3 diagnostic diagram are least affected by an enhanced \heii\ luminosity: completeness fractions of SF-dominated galaxies stay above 50~per cent, and purity fractions of composites do not drop below 25~per cent. To obtain more complete samples of SF-dominated galaxies from observed spectra, we suggest the best UV diagnostic diagrams with \heii\ to be combined with the C43-C32 or C43-OI diagrams without \heii, as the latter result in completeness fractions of SF-dominated galaxies above 80~per cent, as shown in Fig. \ref {UVFrac}. Future progress in the modelling of \heii\ emission in SF galaxies should further improve UV diagnostic diagrams.

\begin{figure*}
\epsfig{file=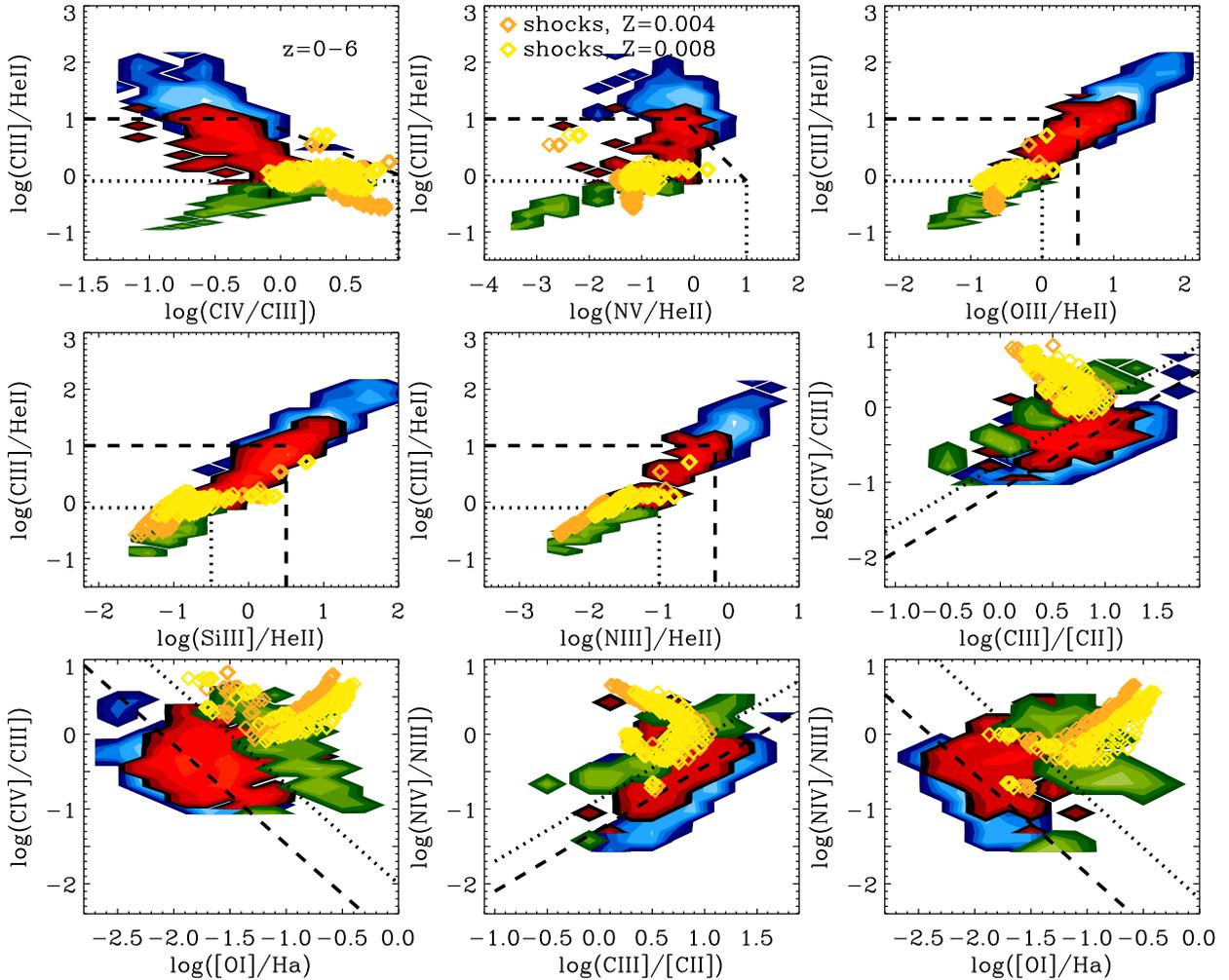,
  width=0.95\textwidth} 
\caption{Radiative shock models at metallicities $Z=0.004$ and 0.008 from \citet[][orange and yellow diamonds, respectively]{Allen08}, plotted over simulations of metal-poor (${\rm N2O2}<-0.8$) SF-dominated (blue 2D histogram), composite (red 2D histogram) and AGN-dominated (green 2D histogram) galaxies at redshifts $z=0$--6, in 9 UV diagnostic diagrams from Section~\ref{UVdiagrams} (different panels, not including EW-based diagrams), together with the selection criteria of Tables~\ref{Selcrit} and \ref{Selcrit_2} (dashed and dotted lines).}\label{UVshocks}    
\end{figure*}

\subsection{UV line ratios from shock excitation}\label{shocks}

Since many observations indicate that galactic outflows from supernovae explosions, stellar winds and AGN exist in a significant fraction of galaxies out to high redshifts \citep[e.g.,][]{Steidel10, Genzel11, Kornei12, Newman12}, shock excitation may be an important contributor to optical and UV-line emission in distant galaxies. Theoretically, shock models were developed by \citet{Dopita03}, \citet[][fast shocks]{Allen08} and \citet[][slow shocks]{Rich10, Rich11}.

As we neglected shock excitation in our analysis so far, in this subsection, we assess (i) which implications shocks can have for our UV selection criteria,\footnote{Note that we do not explicitly discuss the location of shocked regions in optical diagnostic diagrams as many previous works on shock-induced emission lines have already done that.} and (ii) to what extent pure shock-dominated galaxies can be separated from SF-dominated, composite and AGN-dominated galaxies in UV diagnostic diagrams. In the following, we consider the fast-shock models of \citet{Allen08}, calculated with the MAPPINGS III shock and photoionization code. For simplicity, the shock models are not coupled with the properties of simulated galaxies. Instead, we focus on the shock models with lowest two available metallicities, $Z=0.004$ and 0.008 (roughly 1/4 and 1/2 of the present-day solar metallicity; \citealt{Caffau11}) and consider full ranges in all other model parameters (i.e., magnetic-field strength, density of the ionized region, shock velocity and ionization parameter).

In Fig.~\ref{UVshocks}, we plot the $Z=0.004$ and 0.008 shock models of \citet[][orange and yellow diamonds, respectively]{Allen08} over our simulated metal-poor (${\rm N2O2}<-0.8$) SF-dominated (blue 2D histogram), composite (red 2D histogram) and AGN-dominated (green 2D histogram) galaxies at redshifts $z=0$--6 in 9 UV diagnostic diagrams from Section~\ref{UVdiagrams} (different panels, not including EW-based diagrams), together with the selection criteria of Tables~\ref{Selcrit} and \ref{Selcrit_2} (dashed and dotted lines). Pure shock-dominated galaxies are separate from `standard' SF-dominated galaxies, but not from composite and AGN-dominated galaxies. Thus, an observed SF galaxy with strong shocks residing in the composite/AGN region of a UV diagnostic diagram could be mistaken for an active galaxy. Other signatures of nuclear activity (e.g., X-ray, radio emission, integral field spectroscopy) would be required to confirm the presence of an AGN in such a galaxy.

It is worth noting that, in reality, shocks are generally not expected to dominate line emission from a galaxy and are usually associated with SF and AGN activity \citep[see e.g.][]{Kewley13}. Exploring self-consistently the contribution from shocks to the cosmic evolution of emission-line ratios would therefore require that we identify shocks in our simulations and couple their properties (velocity, metallicity) with those of the shock models. We postpone such an investigation to a future study. A further uncertainty consists in the limited metallicity range of the shock models of \citet{Allen08}, not extending below 1/4 solar metallicity. It remains unclear which region of UV diagrams extremely metal-poor shock-dominated galaxies would populate. New shock models will be required to answer this question.


\subsection{Comparison with UV selection criteria proposed in literature}\label{comp}

UV emission-line-based criteria to discriminate between ionizing sources in galaxies have been proposed in a few previous theoretical studies \citep{Feltre16, Nakajima18}, but out of any cosmological context and using separate photoionization models for young stars and AGN, thus, without considering composite galaxies. We now replace our model predictions on UV selection criteria in the context of these previous studies.

Our results agree generally well with those of \citet{Feltre16}, who also identified the C3-C43, C3-N5, C3-O3, C3-Si3 and C3-N3 UV diagrams as powerful diagnostic tools to discriminate between ionizing sources in galaxies.\footnote{Note that, unlike \citet{Feltre16}, we discarded diagnostic diagrams involving neon UV lines, which will be hard to detect because of typically low EW.} This general agreement, at least for metal-poor galaxies, is not surprising, as the nebular-emission models used in our analysis are further developments of those used by \citet[][see Section~\ref{cloudymodels}]{Feltre16}. Interestingly, the revised (harder) spectra of Wolf-Rayet stars in the stellar population models imply that the \heii-based UV diagnostic diagrams of Section~\ref{UVwithHe} are not suitable any longer to differentiate ionizing sources in {\em metal-rich} galaxies (in addition to metal-poor galaxies), as was the case in \citet{Feltre16}. This is because metal-rich SF galaxies now overlap  with composite and AGN-dominated galaxies in these diagrams.

Our results are also largely consistent with those of \citet{Nakajima18}, who provide selection criteria to separate SF- from AGN-dominated galaxies, based on the EW-C3, EW-C4 and C43-C4C3 [i.e., \civ/\ciii\ versus (\civ+\ciii)/\heii] diagnostic diagrams. We chose not to display here the C43-C4C3 diagram, as we found the C3-C4 diagram (based on the same emission lines) to better discriminate between different sources of ionizing radiation. We also note that the EW of AGN models in \citet{Nakajima18} are typically lower than those presented in Section~\ref{UVdiagrams}, as these authors included unattenuated AGN radiation when computing the continuum level (which should \textit{not} be appropriate for type-2 AGN; see Section~\ref{eqwidth}).

Finally, it is of interest to note that all UV diagnostic diagrams proposed by \citet{Feltre16} and \citet{Nakajima18} include the \heii\ line, which can be difficult to reproduce in models of metal-poor galaxies (Section~\ref{HeII}). In the framework of our analysis, we made {\em a first attempt to identify \heii-free diagnostic diagrams} to discriminate between ionizing sources in distant galaxies (Section~\ref{UVnoHe}).

\section{Summary}\label{summary} 

This paper is the second in a series dedicated to the exploration of optical and UV emission-line diagnostic diagrams to classify the nature of ionizing radiation in galaxies over cosmic time. We computed synthetic optical and UV emission lines of galaxies in a cosmological framework by coupling -- in post-processing -- newly developed spectral evolution models, based on photoionization calculations, with a set of 20 high-resolution cosmological zoom-in simulations of massive galaxies, as described in \citet{Hirschmann17}. The latter were performed with the code SPHGal, a modified version of \textsc{Gadget3}, including sophisticated prescriptions for star formation, chemical enrichment \citep{Aumer13}, stellar feedback \citep{Nunez17}, BH growth and AGN feedback \citep{Choi16}. We included nebular emission from young stars \citep{Gutkin16}, AGN \citep{Feltre16} and post-AGB stars \citep{Hirschmann17}. We adopted direct predictions from our simulations for the redshift evolution of global and central interstellar metallicities, C/O abundance ratio, SFR, BH accretion rate, global and central average gas densities, and the age and metallicity of post-AGB stellar populations. Based on these, we selected SF, AGN and post-AGB nebular-emission models for each galaxy and its most massive progenitor at any redshift. By default, we sampled different, reasonable values of undetermined parameters, such as the dust-to-metal mass ratio, ionized-gas hydrogen density and power-law index of AGN ionizing radiation.\\

We can summarize our main results as follows:

\begin{enumerate}

\item The synthetic \oiiihb\ and \niiha\ emission-line ratios of present-day SF, composite and active galaxies predicted by our simulations are in fairly good agreement with those suggested by standard optical selection criteria routinely used to classify the ionizing sources of galaxies in the local Universe.

\item Toward high redshift, standard optical selection criteria in the \oiiihb-\niiha\ diagram can distinguish active from inactive galaxies only in metal-rich galaxies (i.e., with metallicities  $\ga 0.5\,Z_\odot$), as metal-poor galaxies of different types overlap in this diagram. Metal-rich galaxies can be pre-selected observationally, for example using the $\nii/\oii$ or $\niii/\oiiiuv$ ratios. Even so, a differentiation between composite and AGN-dominated galaxies is not feasible at high redshift.

\item To robustly classify the ionizing radiation of metal-poor galaxies, which dominate in the early Universe, we confirm 3 previous diagnostic diagrams from \citet{Nakajima18} and propose 11 novel diagnostic diagrams, all based on luminosity ratios and equivalent widths of emission lines in the {\em UV regime} (the corresponding selection criteria are listed in Tables~\ref{Selcrit} and \ref{Selcrit_2}). UV emission lines tend to become prominent at low metallicities and are accessible to near-infrared spectrographs out to higher redshifts than optical lines. We identify the EW(\ciii) versus \ciii/\heii, EW(\siliii) versus \siliii/\heii\ and \ciii/\heii\ versus \oiiiuv/\heii\ diagrams as most powerful diagnostics to differentiate the ionizing sources in metal-poor galaxies over cosmic time, with completeness and purity fractions (defined in Section~\ref{opticalfractions}) always above $\sim$70~per cent.

\item Current nebular emission models appear to sometimes dramatically underestimate the \heii-line luminosity in very metal-poor galaxies.  We tested the robustness of our UV selection criteria to these uncertainties by exploring the impact of artificially increasing the predicted \heii-line luminosities of our simulated galaxies. We find that SF galaxies typically move toward the regions populated by composites in UV diagnostic diagrams, thereby reducing the completeness fractions of SF galaxies and the purity fractions of composites. The EW(\oiiiuv) versus \oiiiuv/\heii\ diagram is least affected by these changes. In addition, we propose alternative UV diagnostic diagrams {\em not including} the \heii\ line, such as that defined by the \civ/\ciii\ and \ciii/\cii\ ratios, which can also be used in combination with \heii-based diagnostics.

\item When exploring the distribution of pure-shock models in our UV diagnostic diagrams, we find that these tend to fall in regions occupied by composites and AGN galaxies. Thus, an observed SF galaxy with strong shocks residing in the composite/AGN region of a UV diagnostic diagram could be mistaken for an active galaxy. 
\end{enumerate}

The theoretical results presented in this paper provide useful insights into the identification of ionizing sources in metal-poor (primeval) galaxies with metallicities below half solar. Nevertheless, it is important to keep in mind the sparse statistics of our sample of 20 simulated massive galaxies and their main progenitors, and the neglect of galaxies with present-day low masses, which may affect the derived UV selection criteria at any cosmic epoch. The self-consistent relative contribution by radiative shocks to UV- and optical-line ratios, which we neglected in this study, must also be quantified in detail in future work.

In follow-up studies, we plan to identify ultraviolet-line diagnostics most sensitive to the escape of hydrogen-ionizing photons from galaxies, to provide constraints on the re-ionization processes of neutral hydrogen in the early Universe. In addition, we plan to explore the contribution by different ionizing sources to nebular emission in {\em different regions} of a galaxy, i.e. producing spatially resolved emission-line maps to improve the interpretation of modern integral-field spectroscopic observations in terms of galaxy physical parameters.

\section*{Acknowledgements}
We thank Emma Curtis-Lake, and the entire NEOGAL  team  for fruitful discussions. MH, SC and AF acknowledge financial support from the European Research Council (ERC) via an Advanced  Grant under grant agreement no.\,321323--NEOGAL.  AF acknowledges support from the ERC via an Advanced Grant under grant  agreement  no.\,339659--MUSICOS. TN acknowledges support from the DFG priority  program 1573 `Physics of the interstellar medium' from the DFG  Cluster of Excellence `Origin and structure of the Universe'. RSS  is grateful for the generous support of the Downsbrough family, and acknowledges support from the Simons Foundation  through a Simons  Investigator grant.

\bibliographystyle{mn2e}
\bibliography{Literaturdatenbank}

\label{lastpage}

\end{document}